\documentclass[a4paper,fleqn,usenatbib]{mnras}

\usepackage{newtxtext,newtxmath}

\usepackage[T1]{fontenc}
\usepackage{ae,aecompl}
\usepackage{threeparttable}

\usepackage{graphicx}	
\usepackage{amsmath}	
\usepackage{scrextend}
\usepackage{soul}

\title[GRB 141220A]{Coherence scale of magnetic fields generated in early-time
forward shocks of GRBs}

\author[N. Jordana-Mitjans]{
N. Jordana-Mitjans,$^{1}$\thanks{E-mail: N.Jordana@bath.ac.uk}
C. G. Mundell,$^{1}$
R. J. Smith,$^{2}$
C. Guidorzi,$^{3,4,5}$
M. Marongiu,$^{6}$
\newauthor
S. Kobayashi,$^{2}$
A. Gomboc,$^{7}$
M. Shrestha$^{2}$
and I. A. Steele$^{2}$\\
$^{1}$Department of Physics, University of Bath, Claverton Down, Bath, BA2 7AY, UK\\
$^{2}$Astrophysics Research Institute, Liverpool John Moores University, 146 Brownlow Hill, Liverpool, L3 5RF, UK\\
$^{3}$Department of Physics and Earth Science, University of Ferrara, via Saragat 1, I-44122, Ferrara, Italy\\
$^{4}$INFN -- Sezione di Ferrara, Via Saragat 1, 44122 Ferrara, Italy\\
$^{5}$INAF -- Osservatorio di Astrofisica e Scienza dello Spazio di Bologna, Via Piero Gobetti 101, 40129 Bologna, Italy\\
$^{6}$INAF -- Osservatorio Astronomico di Cagliari - via della Scienza 5 - I-09047 Selargius, Italy\\
$^{7}$Center for Astrophysics and Cosmology, University of Nova Gorica, Vipavska 13, 5000 Nova Gorica, Slovenia}

\date{}

\pubyear{2021}

\begin{document}
\label{firstpage}
\pagerange{\pageref{firstpage}--\pageref{lastpage}}
\maketitle

\begin{abstract}
We report the earliest-ever detection of optical polarization from a GRB forward shock (GRB 141220A), measured $129.5-204.3\,$s after the burst using the multi-colour RINGO3 optical polarimeter on the 2-m fully autonomous robotic Liverpool Telescope. The temporal decay gradient of the optical light curves from $86\,$s to $\sim 2200\,$s post-burst is typical of classical forward shocks with $\alpha = 1.091 \pm 0.008$. The low optical polarization $P_{BV} = 2.8 _{- 1.6} ^{+ 2.0} \, \%$ (2$\sigma$) at mean time $\sim 168\,$s post-burst is compatible with being induced by the host galaxy dust ($A_{V, {\rm HG}}= 0.71 \pm 0.15 \,$mag), leaving low polarization intrinsic to the GRB emission itself ---as theoretically predicted for forward shocks and consistent with previous detections of low degrees of optical polarization in GRB afterglows observed hours to days after the burst. The current sample of early-time polarization data from forward shocks suggests polarization from (a) the Galactic and host galaxy dust properties (i.e. $P \sim 1\%-3\%$), (b) contribution from a polarized reverse shock (GRB deceleration time, jet magnetization) or (c) forward shock intrinsic polarization (i.e. $P \leq 2\%$), which depends on the magnetic field coherence length scale and the size of the observable emitting region (burst energetics, circumburst density).
\end{abstract}




\section{Introduction} \label{sec:intro}

Gamma-ray bursts (GRBs) are the brightest flashes of $\gamma$-ray emission in the Universe. After the collapse of a massive star or coalescence of two compact stellar objects \citep{1993ApJ...405..273W,2014ARA&A..52...43B,2017PhRvL.119p1101A,2017ApJ...848L..12A}, the accretion into a new-born compact object powers ultrarelativistic jetted emission that ---via internal dissipation processes--- produces the highly variable and characteristic $\gamma$-ray prompt emission. In the fireball model framework, the ejecta is later on decelerated by the circumburst medium by a pair of external shocks ---a reverse shock propagating back into the ejecta \citep{1992MNRAS.258P..41R,1999ApJ...520..641S, 2000ApJ...545..807K} and a forward shock propagating into the ambient medium--- producing a long-lived afterglow that can be seen from X-rays to radio frequencies (e.g., \citealt{1999PhR...314..575P,2002ARA&A..40..137M, 2004RvMP...76.1143P}).

During the first hundreds of seconds after the burst ---for prompt and reverse shock emission--- different polarimetric signatures are expected for competing jet models: unpolarized emission from weak, tangled magnetic fields for a baryonic jet \citep{1994ApJ...430L..93R,1999ApJ...526..697M} or highly polarized emission due to the presence of globally ordered magnetic fields \citep{2003ApJ...594L..83G, 2003ApJ...597..998L, 2005ApJ...628..315Z,2009MNRAS.394.1182K}. Currently, the discrepancy among the results of time-integrated $\gamma$-ray polarization studies of the prompt emission remains debated \citep{2003Natur.423..415C,2004MNRAS.350.1288R,2005A&A...439..245W,2019ApJ...884..123C,2020A&A...644A.124K,2021MNRAS.504.1939G}. Additionally, GRB 170114A time-resolved analysis suggests the evolution of the polarization degree and angle over a single pulse, reaching values of $P\sim30\%$ \citep{2019A&A...627A.105B}. Early-time optical polarimetric studies of reverse shocks favour a mildly magnetized jet with primordial magnetic fields advected from the central engine (e.g., GRB 090102; \citealt{2009Natur.462..767S}, GRB 120308A; \citealt{2013Natur.504..119M}). However, recent observations suggest that very energetic GRBs can be launched highly magnetized (GRB 190114C: \citealt{2020ApJ...892...97J}).

Whilst polarimetry of the prompt emission and early afterglow determines the jet physics, polarimetric observations of forward shocks allow to test particle acceleration mechanisms (e.g., shock formation, magnetic turbulence), study the dust properties of GRBs' environments and resolve the large-scale geometry of jets at cosmological distances \citep{2003A&A...410..823L,2004MNRAS.354...86R}. In the afterglow framework, the forward shock is powered by shocked ambient medium and tangled magnetic fields are locally generated in shocks and amplified by plasma instabilities (e.g., via Weibel instability; \citealt{1959PhRvL...2...83W,2003ApJ...595..555N,2003ApJ...596L.121S,2005ApJ...618L..75M}). The magnetic field is randomly oriented in space with length scales much smaller than the size of the observable region of the shock. Consequently, the emission is expected to be intrinsically unpolarized when the jet is on-axis \citep{1999ApJ...526..697M}. However, it can be significantly polarized if the random field
is anisotropic (e.g. the averaged strength in the shock normal direction is stronger or weaker)
and the line-of-sight runs almost along the jet edge. The second condition is satisfied
when a jet break is observed \citep{1999ApJ...524L..43S, 1999MNRAS.309L...7G,2004MNRAS.354...86R}. Additionally, the emission can show few percents of polarization due to differential dust extinction \citep{1949Sci...109..165H,2007JQSRT.106..225L} along the Galactic line-of-sight and in the GRB environment (e.g., \citealt{2003A&A...410..823L,2004AandA...420..899K}).

Pioneering polarization studies of GRB afterglows found $P \sim 1\%-3\%$ hours to days after the onset of the GRB in the optical and near-infrared bands (e.g., \citealt{2004ASPC..312..169C}) ---during the forward shock decay. The first polarization constraint was $P<2.3\%$ at $18\,$h post-burst for GRB 990123 \citep{1999Sci...283.2073H}. Observations of GRB 990510 registered steady levels of polarization at $P \sim 1.7 \%$ level from $18\,$h to $21\,$h post-burst, hence validating synchrotron as the emitting mechanism \citep{1999ApJ...523L..33W, 1999A&A...348L...1C}. The polarimetric monitoring of GRB 030329 afterglow $\sim 0.5-38 \, $days post-burst measured low levels with significant variability around the light curve jet break ($P=0.3\%-2.5\%$; \citealt{2003Natur.426..157G}) ---expected for beamed ejecta. Polarization also behaved similarly during the achromatic break of GRB 091018 \citep{2012MNRAS.426....2W}. Recent spectropolarimetric observations of GRB 191221B also detected low levels $P = 1.5\% \pm 0.5\%$ for forward shock emission $\sim 2.9\,$h after the burst \citep{2020arXiv200914081B}.

With the advance of robotic telescopes, the first early-time polarization constraint of an optical afterglow was made $203\,$s post-burst ---during the afterglow onset of GRB 060418. The $P<8\%$ (2$\sigma$) suggested a significant contribution of unpolarized forward shock photons \citep{2007Sci...315.1822M}. GRB 091208B emission was interpreted as forward shock; the $P=10.4\% \pm 2.5\%$ detection during $149-706\,$s post-burst disfavoured the afterglow model and proposed other mechanisms for the amplification of the magnetic field at the front shock, e.g., magnetohydrodynamic instabilities \citep{2012ApJ...752L...6U}. GRB 131030A polarization degree remained steady at $P=2.1\% \pm 1.6\%$ level during $0.18-1.85\,$h post-burst and was associated to dust-induced polarization from the Galactic interstellar medium (ISM; \citealt{2014MNRAS.445L.114K}).

GRB 141220A optical light curve, presented in this paper, resembles those of late-time classical forward shocks detected hours to days after the prompt. In contrast, Liverpool Telescope (LT; \citealt{2004SPIE.5489..679S}) observations indicate that the forward shock was dominating the total emission as soon as $\sim 86 \,$s after the burst. The afterglow was bright at the time of polarization observations --- starting $129.5\,$s post-burst--- hence providing good early-time constraints. GRB 141220A allows checking whether the early-time emission from forward shocks is intrinsically unpolarized as predicted in the fireball model and observed during late-time afterglows.

This paper is structured as follows. In Section \ref{sec:observations}, we detail the data analysis of GRB 141220A follow-up observations by the LT. In Section \ref{sec:Results}, we study the emission decay, the broad-band spectral properties of the burst and the polarization. In Section \ref{sec:discussion}, we check closure relations for GRB 141220A, we discuss which mechanisms can explain the observed polarization levels and we compare GRB 141220A to other early-time forward shock measurements. In Section \ref{sec:conclusions}, we summarize our findings. Throughout this paper, we assume flat $\Lambda$CDM cosmology $\Omega_m = 0.32$, $\Omega_{\Lambda} = 0.68$ and $h=0.67$, as reported by \cite{2020A&A...641A...6P}. We adopt the convention $F_{\nu} \propto t ^{-\alpha} \nu ^{-\beta} $, where $\alpha$ is the temporal index and $\beta$ is the spectral index. The spectral index is related to the photon index like $\beta=\beta_{\rm PI}-1$. Uncertainties are quoted at $1\sigma$ confidence level unless stated otherwise.

\section{Observations and Data Reduction} \label{sec:observations}

\begin{figure*}
	\includegraphics[width=\textwidth]{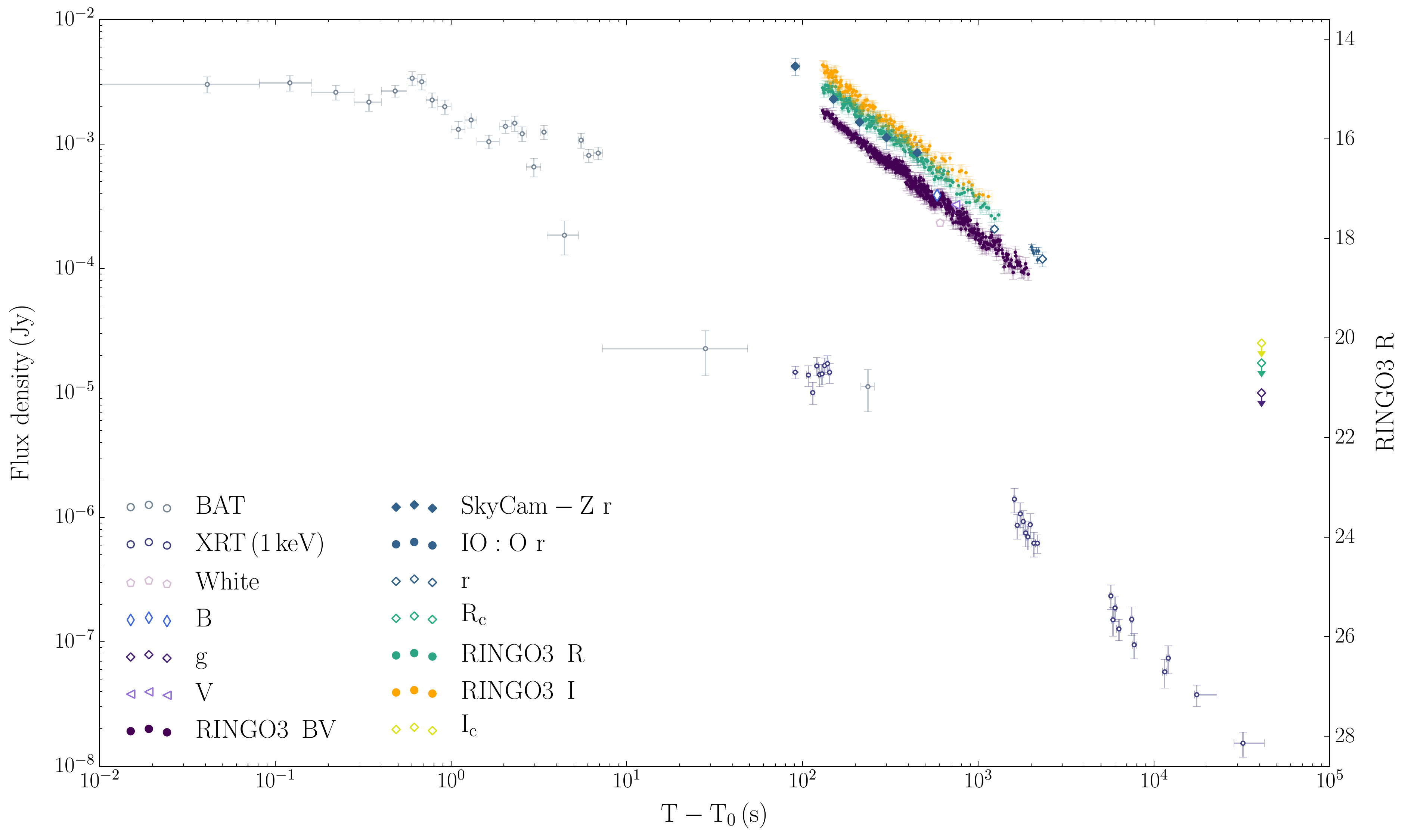}
    \caption{The GRB 141220A multi-wavelength light curves with {\it Swift} BAT, {\it Swift} XRT, LT SkyCam-Z {\it r-}equivalent band, LT RINGO3 {\it BV/R/I} bands and LT IO:O {\it r} band. {\it Swift} observations are obtained from the web interface provided by Leicester University \citep{2009MNRAS.397.1177E}: BAT data are binned to S/N=7 and the absorbed $0.3-10 \,$keV XRT light curve is converted to flux density at $1 \,$keV. For completeness, we include the UV/optical observations reported in GCNs from UVOT \citep{2014GCN.17201....1M}, GTC \citep{2014GCN.17200....1D} and the MITSuME Akeno upper limits \citep{2014GCN.17215....1Y}. The GCN observations do not include filter corrections. In the {\it x-}axis, T$_0$ corresponds to BAT trigger time; in the {\it y-}axis, the flux density is converted to RINGO3 {\it R} magnitude.}
    \label{fig:LC_GRB141220A}
\end{figure*}

On 2014 December 20 at T$_0=\,$06:02:52.7 UT, the {\it Swift} Burst Alert Telescope (BAT;  \citealt{2014GCN.17196....1C}) was triggered by the detection of a pulse of $\gamma$-rays corresponding to the candidate GRB 141220A. The pulses lasted from T$_0-0.9\,$s to T$_0+7.3\,$s and consisted of one structured peak ---centred at $\sim $T$_0+0.5\,$s--- followed by two fainter ones at $\sim {\rm T}_0+3.5\,$s and $\sim {\rm T}_0+6.0\,$s. The emission was detectable until $\sim 30 \,$s post-burst (see Fig.~\ref{fig:LC_GRB141220A}; \citealt{2009MNRAS.397.1177E}). 

In the $15-350\,$keV band, BAT measured a peak energy $E_{\rm peak} =117 \pm 45 \,$keV with the $90\%$ of the burst fluence released during T$_{90} = 7.2 \pm 0.5\,$s \citep{2014GCN.17202....1S}. Both {\it Fermi} Gamma-ray Burst Monitor (GBM; \citealt{2014GCN.17205....1Y}) and Konus{\it -Wind} (KW; \citealt{2014GCN.17207....1G}) detected the $\gamma$-ray prompt with similar T$_{90}$ in the $50-300\,$keV and $20\, \rm{keV}-10\,$MeV energy range, respectively. KW detected the event up to $\sim 4 \,$MeV and measured a peak energy $E_{\rm peak} =139 ^{+10} _{-9}\,$keV, isotropic energy $E_{\rm iso}=(2.29\pm 0.12)\times10^{52}\,$erg and a peak isotropic luminosity $L_{\rm iso}=(2.89 \pm 0.04)\times10^{52}\,$erg$\,$s$^{-1}$ \citep{2017ApJ...850..161T}. Given the duration of the burst and the hardness ratio\footnote{The ratio between the fluences in the $50-100 \,$keV and $25-50 \,$keV energy bands.} $S(50-100 \, {\rm keV})/S(25-50 \, {\rm keV}) \sim 1.6$ \citep{2016ApJ...829....7L}, GRB 141220A is classified as a long-soft GRB type \citep{1993ApJ...413L.101K,2013ApJ...764..179B, 2016ApJ...829....7L} ---typically associated with a collapsar origin. This classification is further supported by $E_{\rm peak}/{\rm T_{90}} \sim 16\,$keV$\,$s$^{-1}$ ---the ratio is $E_{\rm peak}/{\rm T_{90}} \lesssim 50\,$keV$\,$s$^{-1}$ for $99\%$ of long GRBs in BATSE sample \citep{2015MNRAS.451..126S}.

A bright optical afterglow of $ 14.84 \pm 0.17 \,$mag was detected $86\,$s after BAT trigger at the GRB 141220A location by the 0.2-m SkyCam-Z telescope (see Section \ref{sec:SkyCam}) ---attached to the LT.

At $87.2\,$s after the burst, the {\it Swift} X-ray Telescope (XRT; \citealt{2014GCN.17203....1G}) also started observing the field. The XRT light curve has two distinct segments separated by a gap in observations: the first $57\,$s were done in Windowed Timing (WT) mode, whilst the steepest decay was observed in Photon Counting (PC) mode (see Fig.~\ref{fig:LC_GRB141220A}; \citealt{2009MNRAS.397.1177E}). 

The 2-m LT reacted automatically to Swift alert \citep{2006PASP..118..288G} with the three-band optical polarimeter and imager RINGO3 with observations starting at $129.5\,$s post-burst (see Section \ref{sec:RINGO3}). LT observations consisted of $3\times 10 \,$minute epochs of RINGO3 instrument, followed by $ 6 \times 10 \,$s frames with the {\it r} band of the Optical Wide Field Camera (IO:O; see Section \ref{sec:IOO}) and $7\times 10 \, $minute more with RINGO3.

At $\sim 607\,$s post-burst, the 0.3-m {\it Swift} Ultraviolet/Optical Telescope (UVOT; \citealt{2014GCN.17201....1M}) detected an optical counterpart of $17.30 \pm 0.07\,$mag in the white band. At $\sim 20.6\,$minutes post-burst, the 10.4-m Gran Telescopio Canarias (GTC; \citealt{2014GCN.17200....1D, 2014GCN.17198....1D}) detected GRB 141220A with $18.11 \pm 0.05 \,$mag in the {\it r} band and derived a spectral redshift of ${\rm z} =1.3195$.

\subsection{Calibration of RINGO3 {\it BV/R/I} Band Observations} \label{sec:RINGO3}

The RINGO3 instrument employs a spinning polaroid and two dichroic that split the light beam into three optical bands \citep{2012SPIE.8446E..2JA}. From $129.5\,$s post-burst, the source intensity was sampled every $1.15\,$s at eight polaroid angles by three fast-readout cameras. These $1.15 \,$s exposures were automatically co-added by the telescope pipeline\footnote{\url{https://telescope.livjm.ac.uk/TelInst/Pipelines/}} into $10 \times 1 \, $minute integrations. Given RINGO3 instrumental configuration, the photometry is derived integrating the intensity of all eight frames (see Section \ref{sec:LC_reduc}) and the polarization degree and angle, measuring the relative intensity of the source at each of the eight polaroid positions (see Section \ref{sec:P_reduc}). For both analyses, we use aperture photometry to derive the source flux with Astropy Photutils package \citep{2016ascl.soft09011B}.

\subsubsection{Photometric Calibration of Optical Light Curves} \label{sec:LC_reduc}

The signal-to-noise ratio (S/N) of the optical transient (OT) was high at the start of observations; the OT was detected at S/N$\sim30$ in each of the first $ \sim 25 \times 1.15 \,$s frames of the {\it BV} band. Given its brightness, we use the $1.15\,$s exposures instead of the $1 \,$minute integrations. We follow the same procedure as in \cite{2020ApJ...892...97J} and we dynamically co-add the $1.15\,$s frames such that the OT reaches a minimum threshold of ${\rm S/N}=10$, which gives a $\sim 0.4-0.1 \,$mag photometric precision. {\it BV} band observations after $\sim 550 \, $s post-burst are the result of integrating frames; the {\it R/I} bands have lower S/N and the frame co-adding starts at $\sim $T$_0 + 200\, $s. The {\it BV} photometry after $\sim$T$_0 + 2000 \,$s is discarded because the OT fades beyond the S/N threshold; the source S/N is lower in the {\it R/I} bands and we only accept photometry up to $\sim 1350 \,$s and $\sim 1200 \,$s post-burst, respectively. Integrations of lower S/N thresholds do not present statistically significant structure (within 3$\sigma$) in addition to the OT constant emission decay; frame binnings with thresholds S/N$\, \gg 10$ discard late-time photometry. We check the stability of RINGO3 during the OT observations with the stars in the field: a $11.5\,$mag star and three $15-19\,$mag stars. Using the OT binning for the photometry, the stars present a $\sim 0.05 \,$mag deviation from the mean.

We calibrate the photometry in absolute flux in RINGO3 {\it BV/R/I} bandpass system following \cite{2020ApJ...892...97J}. Observations of three dereddened A0 type stars (HD 87112, HD 92573, HD 96781; \citealt{2000A&A...355L..27H}), including the GRB 141220A field, were scheduled via LT phase2UI\footnote{\url{http://telescope.livjm.ac.uk/PropInst/Phase2/}} using the same instrumental setup of the night of the burst. The observations were automatically dispatched during the nights of 2018 May 2-3. We correct for the mean Galactic extinction ($E_{\rm BV,MW} = 0.0128 \pm 0.0005$ is derived from a $5' \times 5'$ field statistic\footnote{\url{https://irsa.ipac.caltech.edu/applications/DUST/}}; \citealt{1998ApJ...500..525S}) and we use the RINGO3 magnitude-to-flux conversion from \cite{2020ApJ...892...97J}. This calibration adds $\sim 0.05 \,$mag uncertainty to the photometry. 

In Table~\ref{tab:phot} and Fig.~\ref{fig:LC_GRB141220A}, we present the photometry of GRB 141220A in RINGO3 {\it BV/R/I} bands. Note that magnitudes and fluxes are not corrected for the host galaxy extinction (see Section \ref{sec:spectra_sed}).

\subsubsection{Calibration of Optical Polarization}  \label{sec:P_reduc}

We measure the flux of the OT at each of the eight rotor positions of the polaroid and, following \cite{2002A&A...383..360C}, we convert them to the Stokes parameters (q-u). The confidence levels of the Stokes parameters and the polarization degree and angle are determined with a Monte Carlo error propagation starting from $10^6$ simulated flux values for each polaroid position.

\begin{figure}
	\includegraphics[width=\columnwidth]{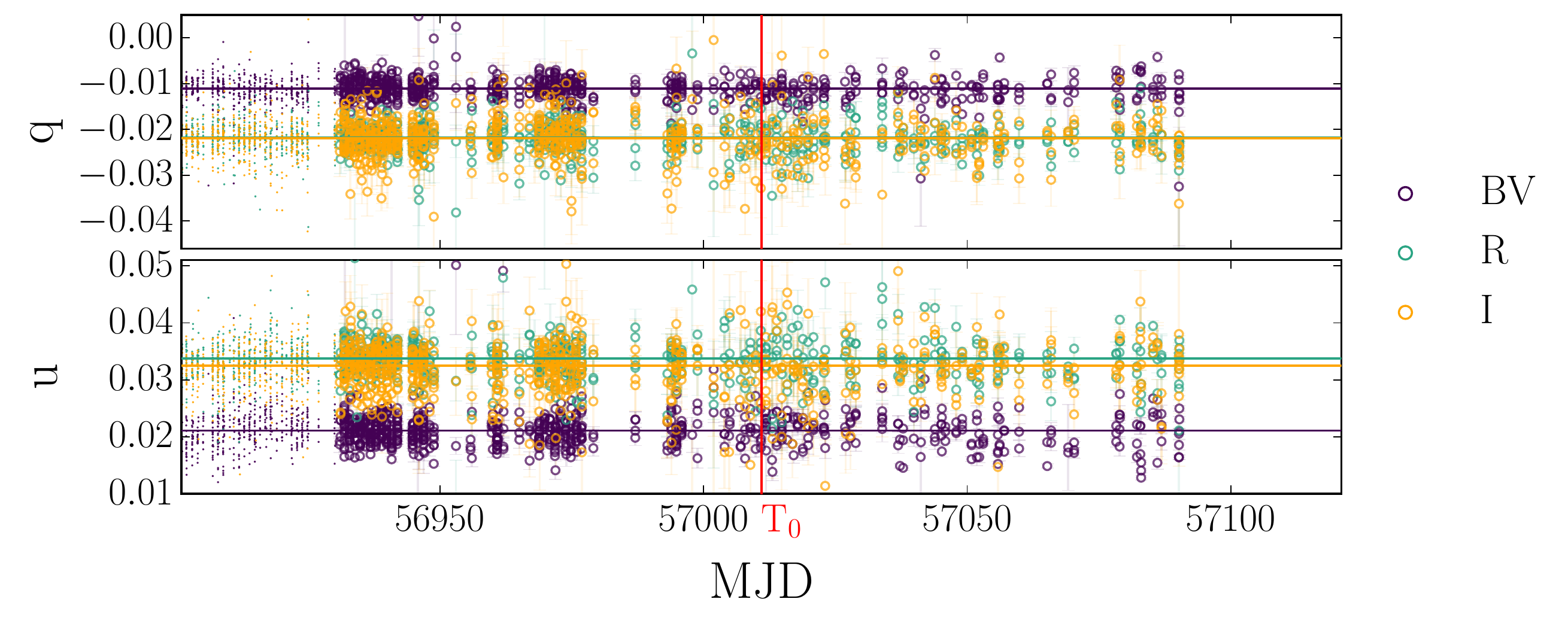}
    \caption{The RINGO3 {\it BV/R/I} Stokes parameters (q-u) of the unpolarized standard stars (BD +32 3739, BD +33 2642, BD +28 4211, HD 14069, HD 109055 and G191-B2B) used for the determination of GRB 141220A instrumental polarization (circles). The red vertical line indicates GRB 141220A burst date (T$_0$) and the coloured horizontal lines correspond to the median q-u value per band.}
    \label{fig:qu_inst}
\end{figure}

RINGO3 instrumental polarization is subtracted in the Stokes parameters plane and the polarization angle is standardized to a reference \citep{2016MNRAS.458..759S}. As RINGO3 is regularly taking observations of standards stars, we choose observations taken during a period of $\pm 80 \,$days on either side of the burst date T$_0$ (see Figure \ref{fig:qu_inst}). From 2014-10-01 to 2015-03-10, the following $9-11\,$mag unpolarized standards stars were observed with RINGO3: BD +32 3739, BD +33 2642, BD +28 4211, HD 14069, HD 109055 and G191-B2B. The instrumental polarization is determined with the median q-u of $\sim 720$ observations of the standards per RINGO3 band. We estimate a $\sim 3 \times 10^{-3}$ precision in the definition of the instrumental q-u, which could introduce no more than $P \sim 0.5\%$ in polarization after their subtraction. We also check for any monotonic trend of the instrumental q-u during the T$_0 \pm 80 \,$days' time-window; the Pearson's correlation coefficients of each band are not significant with $|r|< 0.08$ and $p$-values$\,>0.1$.

\begin{table}
	\centering
	\caption{The GRB 141220A photometry in RINGO3 {\it BV/R/I} bands, SkyCam-Z {\it r-}equivalent band and IO:O {\it r} band.}
	\label{tab:phot}
	\begin{tabular}{ccccccc}
		\hline
		Band & t$_{\rm mid}$ & t$_{\rm exp}/2$  & mag & mag$_{\rm \, err}$  & F$_{\nu}$ & F$_{\nu \, \rm{err}}$\\
		 & (s) & (s) & & & (Jy) & (Jy) \\
		\hline
		{\it BV} &    130.1 &      0.6 &    15.73 &     0.08 & $1.86 \times 10^{-3}$ &  $1.3 \times 10^{-4}$\\
		{\it BV} &    131.3 &      0.6 &    15.76 &     0.08 & $1.81 \times 10^{-3}$ &  $1.3 \times 10^{-4}$\\
		{\it BV} &    132.4 &      0.6 &    15.78 &     0.08 & $1.77 \times 10^{-3}$ &  $1.3 \times 10^{-4}$\\
		{\it BV} &    133.6 &      0.6 &    15.88 &     0.08 & $1.61 \times 10^{-3}$ &  $1.2 \times 10^{-4}$\\
		{\it BV} &    ... &     ...  &    ... &     ... & ... &  ...\\
		\hline
	\end{tabular}
    \begin{tablenotes}
        \item[a] {\it Note}. t$_{\rm mid}$ corresponds to the mean observing time and t$_{\rm exp}$ to the length of the observation window. Magnitudes and flux density values are corrected for atmospheric and Galactic extinction. Table~\ref{tab:phot} is published in its entirety in the machine-readable format. A portion is shown here for guidance regarding its form and content.
    \end{tablenotes}
\end{table}

Following \cite{2013Natur.504..119M} and \cite{2020ApJ...892...97J}, we co-add the $1.15$-s frames of the first $10$-minute epoch so that the OT has maximum S/N. The {\it BV} band provides the highest S/N for the time window  $129.5-204.3\,$s post-burst with polarization $P_{BV}= 2.8  _{-  1.6} ^{+  2.0} \, \%$ at a $2\sigma$ confidence level. The OT is well detected at an average S/N$\sim 60$ in each of the eight polaroid position images. The remaining measurements of the {\it BV} band ---and the {\it R/I} bands--- are derived maximizing the S/N. The polarization angle is standardized to a reference using $\sim 485$ observations of $9-11.5\,$mag polarized standard stars taken during the T$_0 \pm 80 \,$days' time-window: BD +25 727, BD +59 389, HD 155528, VI CYG 12, BD +64 106 and Hiltner 960 \citep{1990AJ.....99.1243T,1992AJ....104.1563S}. GRB 141220A polarization measurements are presented in Table \ref{tab:pol_tab} with the polarization degree and the angle uncertainty quoted at $2\sigma$ confidence level.

We estimate RINGO3 depolarizing factor\footnote{Note that $P/D$ is the true polarization such that an instrument with no depolarization has $D=1$.} ($D$) during the T$_0 \pm 80 \,$days period using the $\sim 485$ observations of the $P=4.5\%-9\, \%$ polarized standard stars and $\sim 95 $ observations of the $P \sim 50\%$ polarized planetary nebula of CRL 2688 \citep{1976ApJ...204L..25S}. After q-u ellipticity corrections \citep{ArnoldThesis}, RINGO3 depolarizing effect is negligible in the {\it BV} band ($D_{\it BV}=1.00\pm 0.01$) and small in the {\it R/I} bands ($D_{\it \lbrace R, I\rbrace}=0.98 \pm 0.01, 0.94 \pm 0.01$). Because {\it R/I} band polarization measurements are dominated by noise (resulting in upper limits), this correction is not applied to the polarization measurements of Table \ref{tab:pol_tab}. We check the stability of RINGO3 instrument during observations using the bright star in the {\it R/I} band field-of-view that has polarization $P_{\lbrace R, I\rbrace} =0.22_{- 0.11} ^{+ 0.13}  \%,  0.23_{- 0.14} ^{+ 0.18} \,  \%$. Using the OT binning, the polarization presents a $P_{ \lbrace R, I \rbrace } \sim 0.07 \%$ deviation from the mean. We test the robustness of the polarization measurements with $1.5-3 \,$FWHM photometric apertures for the OT and the bright star. There is no polarization variations within $1\sigma$. 

\begin{table*}
	\centering
	\caption{The GRB 141220A polarization observations with RINGO3 {\it BV/R/I} bands.}
	\label{tab:pol_tab}
	\begin{tabular}{cccccccccccc}
		\hline
		Band & t$_{\rm mid}$ & t$_{\rm exp}/2$ & S/N & q & q$_{\rm err}$ & u & u$_{\rm err}$ & $P$ & $P_{\rm err} $ & $\theta$ & $\theta_{\rm err}$ \\
		& (s) & (s) &  &  &  &  &   & ($\%$)  & ($\%$) & (\textdegree) & (\textdegree) \\
		\hline
{\it BV} & 166.9 &  37.4 &  60 & -0.017 & 0.009 & 0.023 & 0.009 &   2.8 & $_{-  1.6} ^{+  2.0}$ &  63 & 20 \\
{\it BV} & 253.8 &  49.5 &  42 & 0.006 & 0.014 & 0.014 & 0.014 & $<  4.5$ & - & - & - \\
{\it BV} & 514.4 & 211.1 &  38 & -0.008 & 0.015 & 0.014 & 0.015 & $<  4.9$ & - & - & - \\
{\it BV} & 1043.0 & 298.0 &  21 & -0.009 & 0.026 & 0.008 & 0.026 & $<  7.5$ & - & - & - \\
{\it BV} & 1658.5 & 298.0 &  11 & -0.042 & 0.051 & 0.022 & 0.051 & $< 16.4$ & - & - & - \\
\hline
{\it R} & 166.9 &  37.4 &  30 & -0.006 & 0.018 & -0.005 & 0.018 & $<  5.2$ & - & - & - \\
{\it R} & 282.0 &  77.7 &  23 & -0.004 & 0.024 & -0.005 & 0.024 & $<  6.7$ & - & - & - \\
{\it R} & 542.6 & 182.9 &  17 & 0.034 & 0.033 & 0.020 & 0.033 & $< 11.4$ & - & - & - \\
{\it R} & 1043.0 & 298.0 &  11 & 0.083 & 0.049 & -0.005 & 0.049 & $< 19.2$ & - & - & - \\
{\it R} & 1658.5 & 298.0 &   7 & 0.007 & 0.085 & -0.122 & 0.085 & $< 31.4$ & - & - & - \\
\hline
{\it I} & 166.9 &  37.4 &  23 & -0.022 & 0.024 & 0.016 & 0.024 & $<  8.1$ & - & - & - \\
{\it I} & 282.0 &  77.7 &  17 & -0.016 & 0.033 & 0.010 & 0.033 & $<  9.$8 & - & - & - \\
{\it I} & 542.5 & 182.9 &  14 & -0.007 & 0.040 & 0.061 & 0.040 & $< 14.9$ & - & - & - \\
{\it I} & 1043.0 & 298.0 &   9 & -0.086 & 0.063 & -0.014 & 0.063 & $< 22.8$ & - & - & - \\
{\it I} & 1658.5 & 298.0 &   6 & 0.013 & 0.088 & 0.120 & 0.088 & $< 32.0$ & - & - & - \\
	\end{tabular}
    \begin{tablenotes}
        \item[a] {\it Note}. $P$ and $\theta$ uncertainties and upper limits are quoted at $2\sigma$ confidence level.
    \end{tablenotes}
\end{table*}

\subsection{Photometric Calibration of {\it White}-light SkyCam-Z Observations} \label{sec:SkyCam}

The SkyCam-Z telescope\footnote{\url{https://telescope.livjm.ac.uk/TelInst/Inst/SkyCam/}} is attached to the LT mount and continuously monitors LT field-of-view with a $1^{\circ} \times 1^{\circ}$ coverage in a white band ---taking a single exposure of $10\,$s every minute. The SkyCam-Z started the first 10-s exposure of GRB 141220A field at $86\,$s post-burst, before RINGO3 observations. However, the LT telescope was still slewing towards the target at that time and the first frame has trails; we correct the flux offset of the first frame with the succeeding SkyCam-Z observation (at T$_0+145\,$s) using 55 stars detected in both fields. Observations starting at T$_0+ 265\,$s are co-added to reach a minimum threshold of S/N=5. We standardize the white-band photometry with the {\it r-}band using five $8-12\,$mag stars from the SDSS DR12 catalogue \citep{2015ApJS..219...12A}. In Table~\ref{tab:phot} and Fig.~\ref{fig:LC_GRB141220A}, the photometry is corrected for mean the Galactic extinction ($A_{r} = 0.035 \pm 0.001\, $mag) but not for host galaxy extinction (see Section \ref{sec:spectra_sed}).

\subsection{Photometric Calibration of {\it r}-band IO:O Observations}  \label{sec:IOO}

The IO:O\footnote{\url{https://telescope.livjm.ac.uk/TelInst/Inst/IOO/}} observations started at T$_0 + 2002\,$s with the {\it r} filter. Given that the OT is detected at S/N$\, \sim 100$ per frame, we individually derive the OT flux. We standardize the photometry using ten $12-17\,$mag stars from the SDSS DR12 catalogue \citep{2015ApJS..219...12A}. In Table~\ref{tab:phot} and Fig.~\ref{fig:LC_GRB141220A}, we present the IO:O photometry in the {\it r} band. The IO:O light curve is corrected for the mean Galactic extinction ($A_{r} = 0.035 \pm 0.001\, $mag) but not for host galaxy extinction (see Section \ref{sec:spectra_sed}).

\section{Results} \label{sec:Results}

In this section, we present the time-resolved polarization measured in three optical bands (see Section~\ref{sec:pol}), the optical light curves in three bands and a white band (see Section~\ref{sec:LC}), the spectral evolution in the optical bands (see Section~\ref{sec:spectra_opt}) and the combined optical/X-rays spectral energy distributions (SEDs; see Section~\ref{sec:spectra_sed}).

\subsection{Polarization in Three Optical Bands} \label{sec:pol}

\begin{figure}
	\includegraphics[width=\columnwidth]{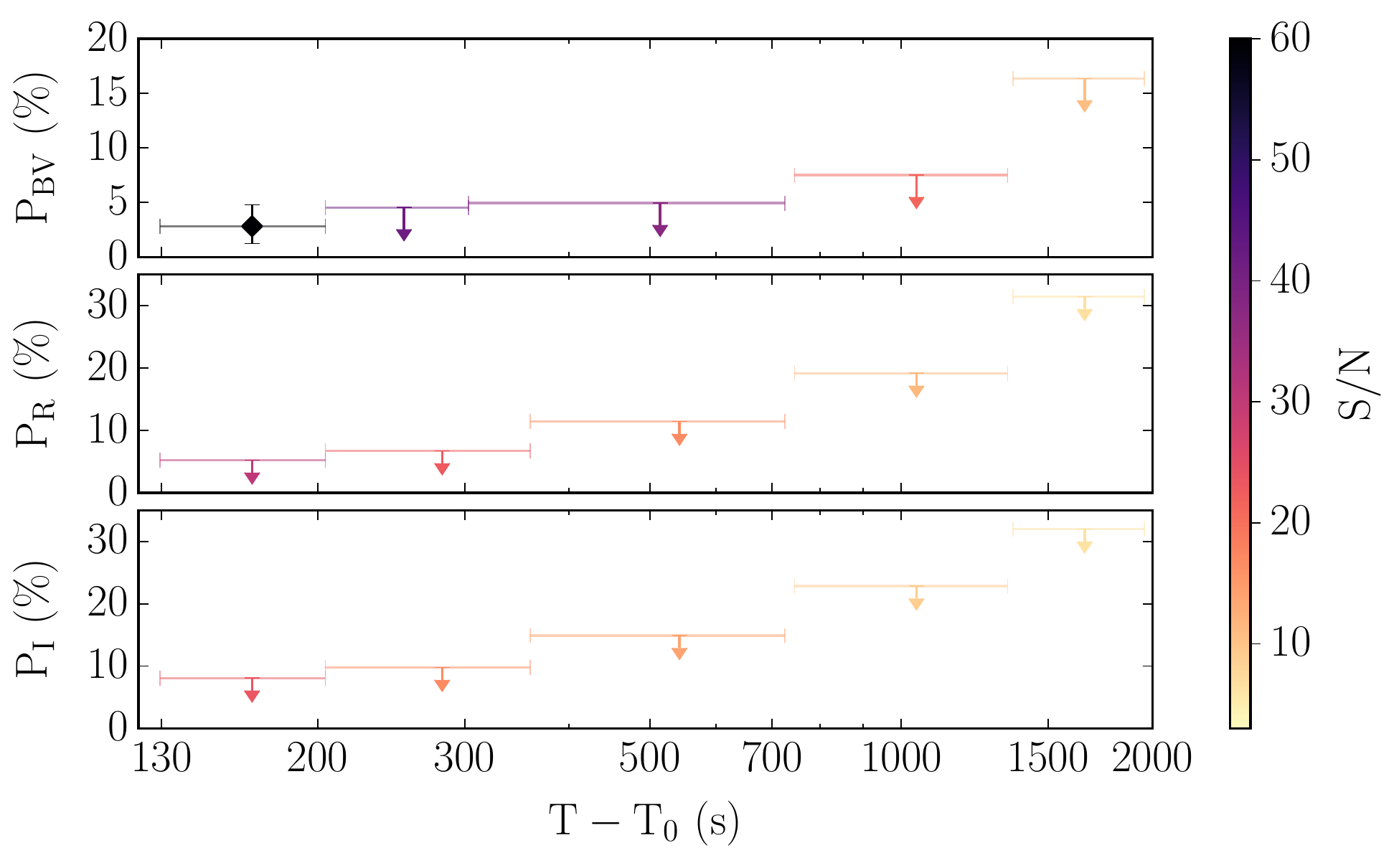}
    \caption{The temporal evolution of GRB 141220A polarization in RINGO3 {\it BV/R/I} bands with a colour map of the OT S/N. In the {\it BV} band, the black marker corresponds to the only polarization measurement.}
    \label{fig:Polarization}
\end{figure}

We detect low polarization in RINGO3 {\it BV/R/I} bands since the start of observations at $129.5\,$s post-burst (see Fig. \ref{fig:Polarization}). The best constraint is the polarization measurement $P_{BV} = 2.8 _{-  1.6} ^{+  2.0} \, \%$ at $2\sigma$ confidence level, with polarization angle $\theta_{BV} =63 $\textdegree$ \pm  20 \,$\textdegree (2$\sigma$). This measurement corresponds to the time window $129.5-204.3\,$s post-burst, i.e. $55.9-88.1\,$s after the burst in the GRB rest-frame. Additionally, we use the permutation test detailed in \citealt{2017ApJ...843..143S} to prove that the observed polarization is a measurement and not the result of random noise in an unpolarized source. Consequently, we randomly reorder the eight flux values, corresponding to each angle position of the polaroid, to create a sample of 40320 polarization values with the same S/N properties of the observed polarization. If the observed polarization lays within the simulated distribution ($p_{\rm t} \lesssim 0.75$), it indicates that there is high probability of being the result of the random noise. The observed polarization $P_{BV} = 2.8 _{-  1.6} ^{+  2.0} \, \%$ belongs to the $p_{\rm t} = 0.97$ upper quantile of the overall distribution; there is only $3\%$ probability that this measurement is consistent with zero polarization.

In the {\it BV} band, the polarization is further constrained at $P_{BV}<5\%$ ($2\sigma$) level until $726\,$s post-burst. Afterwards, the $2\sigma$ upper limits increase significantly with time due to the decreasing brightness of the OT, which cannot be compensated by integrating more frames. The {\it R/I} bands have less S/N and the observed polarization is within the instrument sensitivity, ranking $p_{\rm t} < 0.75$ in the polarization permutation test. The best constraints in {\it R/I} bands are the $2\sigma$ upper limits $P_{\lbrace R, I \rbrace}<5.2\%, 8.1\%$ during $129.5-204.3\,$s post-burst ---consistent with the cotemporal {\it BV} band measurement.

\subsection{Optical Light Curves} \label{sec:LC}

\begin{figure}
	\includegraphics[width=\columnwidth]{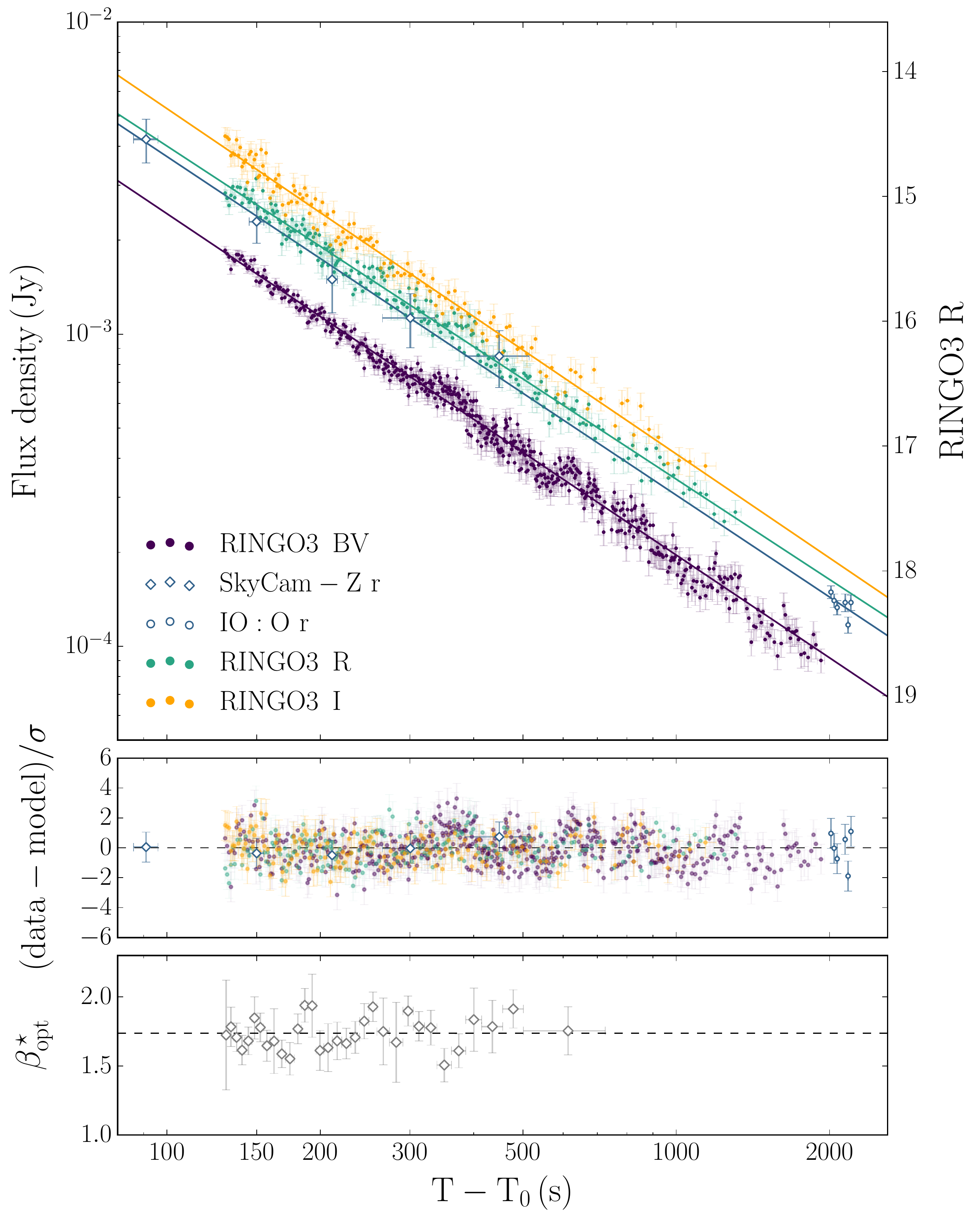}
    \caption{The GRB 141220A {\it BV/r-}equivalent{\it /R/I} band light curves fitted with power laws of common slope $\alpha_{\rm opt}=1.091 \pm 0.008$. In the {\it x-}axis, T$_0$ corresponds to BAT trigger time. In the {\it y-}axis, the flux density is converted to RINGO3 {\it R} magnitude. The middle panel corresponds to the residuals of the fit. The bottom panel shows the evolution of the optical spectral index $\beta _{\rm opt} ^{\star}$; the dashed line is the mean. Note that $\beta _{\rm opt} ^{\star}$ is not corrected by the host galaxy extinction. }
    \label{fig:LC_GRB141220A_fit}
\end{figure}

The {\it BV/R/I} light curves are satisfactorily modelled with power laws that have compatible decay indexes (see Table~\ref{tab:summary_index}). We note that the {\it BV} band has worse $\chi^2$ statistics than the {\it R/I} bands due to smaller photometric uncertainties\footnote{The statistical uncertainty of the {\it BV} band is likely underestimated; for the first $\sim 300 \, $s of the light curve, it is smaller than the $\sim 0.05 \,$mag deviation we measure in field stars.} (higher S/N). The SkyCam-Z {\it White} and IO:O {\it r} band observations are also fitted in a joint {\it r-}equivalent light curve. If we simultaneously fit all light curves with a single power law, the best-fitting model indicates that the emission was decaying as $\alpha _{\rm opt}= 1.091 \pm 0.008$ since $86\,$s and up to $\sim 2200\,$s after the burst (see Fig.~\ref{fig:LC_GRB141220A_fit}).

\begin{table}
	\centering
	\caption{Temporal ($\alpha$) and spectral indexes ($\beta$) of GRB 141220A emission.}
	\label{tab:summary_index}
	\begin{tabular}{ccccc}
		\hline
		Index & Result & $\chi^2$/dof & Fig.\\
		\hline
		$\alpha_{BV}$ & $1.095 \pm 0.005$ & $657.7/535$ &-\\
		$\alpha_{r}$    & $1.09 \pm 0.03$ & $7.4/9$  & - \\
		$\alpha_{R}$   & $1.067 \pm 0.009$ & $203.7/220$ &-\\
		$\alpha_{I}$    & $1.105 \pm 0.013$ & $122.6/136$ & -\\
		$\alpha_{\rm opt}$  & $1.091 \pm 0.008$ & $999.2/903$ & \ref{fig:LC_GRB141220A_fit} \\
		\hline
		$\beta_{\rm xrays}$ $^a$& $1.1 \pm 0.2$ & $21.2/17$ & \ref{fig:SED_broadband} \\
		$\beta_{\rm opt}$ $^a$& $0.6 \pm 0.2$ & - & \ref{fig:SED_broadband} \\
		$\beta_{\rm opt} ^{\star}$ $^b$&  $1.74 \pm 0.11$ & $26.2/31$ & \ref{fig:LC_GRB141220A_fit} \\
		\hline
	\end{tabular}
	
	    \begin{tablenotes}
        \item[a] $^a$ The spectral indexes are linked as $\beta_{\rm xrays} =\beta_{\rm opt} + 0.5$.
         \item[b] $^b$ $\beta_{\rm opt} ^{\star}$ is not corrected for host galaxy extinction; $\beta_{\rm opt} ^{\star}$ is an upper limit of $\beta_{\rm opt}$.
    \end{tablenotes}
\end{table}

\subsection{Spectral Evolution of the Emission} \label{sec:spectra}

\subsubsection{Optical Spectral Index}  \label{sec:spectra_opt}

Given RINGO3 simultaneous three-band imaging, we study the evolution of the optical spectral index. We follow \cite{2020ApJ...892...97J} procedure and we co-add the $1.15$-s frames of the lowest S/N band ({\it I} band) to reach a minimum  threshold of S/N=20, which gives $\sim 0.05-0.07 \,$mag spectral precision through observations and discards photometry after $\sim 700\,$s post-burst. We use the {\it I-}band binning to co-add the {\it BV/R} band frames. We subtract the Galactic extinction ($E_{B-V, \, {\rm MW}} = 0.0128\pm 0.0005$; \citealt{1998ApJ...500..525S}) from the flux-calibrated photometry. For each temporal bin, we fit a power law to the three-band SED and we obtain the spectral index $\beta _{\rm opt}^{\star}$.

The evolution of the optical spectral index is consistent with a mean $\beta _{\rm opt} ^{\star} = 1.74 \pm  0.11$ ($\chi^2$/dof$=26.2/31$ and $p$-value=0.7); the optical emission does not suffer any significant spectral change during $130-725\,$s post-burst (see Fig.~\ref{fig:LC_GRB141220A_fit} bottom panel). Note that the spectral index $\beta _{\rm opt} ^{\star}$ is not corrected by the host galaxy extinction and, therefore, it is an upper limit of the true optical spectral index $\beta_{\rm opt}$.

\subsubsection{Broadband Spectral Energy Distribution}  \label{sec:spectra_sed}

To infer the optical to X-rays synchrotron spectrum, we analyse the broadband SEDs using the time-resolved 0.3-10 keV XRT and RINGO3 observations. We extract the XRT spectra that has cotemporal RINGO3 observations using the web interface provided by Leicester University\footnote{\url{https://www.swift.ac.uk/user_objects/}} based on HEASoft (v. 6.22.1; \citealt{1995ASPC...77..367B}). We then obtain a single RINGO3 photometric measurement per band by co-adding the $1.15\,$s frames that correspond to the XRT time interval. XRT has a gap of observations between WT and PC mode; consequently, the two broadband SEDs correspond to the time intervals $129-148\,$s and $1589-1957\,$s.

\begin{figure}
	\includegraphics[width=\columnwidth]{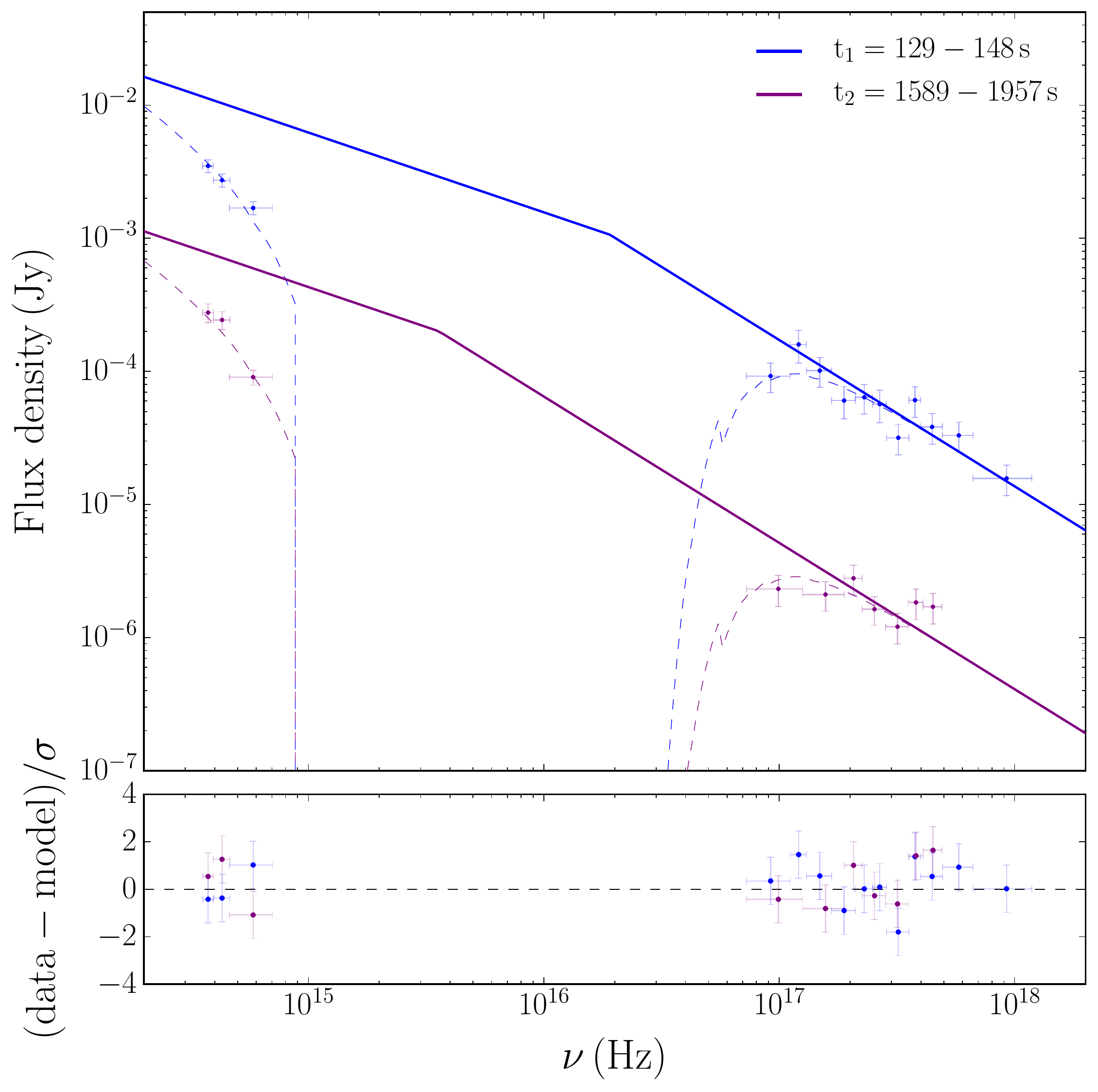}
    \caption{The GRB 141220A broadband SEDs with RINGO3 and XRT observations \citep{2009MNRAS.397.1177E}. The SEDs are best fitted with a broken power law model (solid line) that accounts for extinction in the optical and total hydrogen absorption in the X-rays (dashed line). The results of the fit are: $\beta_{\rm opt}=0.6 \pm 0.2$, $\beta_{\rm xrays}=1.1 \pm 0.2$, $E_{\rm break, 1} = 0.08 ^{+0.04} _{-0.03} \,$keV,  $E_{\rm break, 2} = 0.015 \pm 0.005\,$keV and, in the host galaxy rest-frame, $A_{V, \, {\rm HG}} =  0.71 \pm 0.15\, $mag and $N_{\rm H, \, HG} = (3 \pm 2) \times 10^{21}\,$cm$^{-2}$ ($\chi^2/$dof$=21.2/17$ and $p$-value$=0.2$). The error bars in the optical frequencies are the FWHM of RINGO3 {\it BV/R/I} bandpasses. The bottom panel corresponds to the residuals of the fit.}
    \label{fig:SED_broadband}
\end{figure}

We model the combined optical/X-ray data with XSPEC (v. 12.9.1; \citealt{1999ascl.soft10005A}) using $\chi^2$ statistics. To ensure the Gaussian limit, we use the GRPPHA tool to group the XRT spectra such that there are at least 20 counts per bin. The models we employ are either power laws (powerlaw in XSPEC) or connected power laws (bknpower) that account for the rest-framed host galaxy and MW total hydrogen column absorption and dust extinction (phabs*zphabs*zdust*zdust). For both SEDs, we fix the MW dust extinction to $E_{B-V, \, {\rm MW}} = 0.0128\pm 0.0005$ \citep{1998ApJ...500..525S}, the total hydrogen column absorption to $N_{\rm H, \, MW}= 1.33 \times 10^{20}\,$cm$^{-2}$ (derived using the NHtot Swift tool\footnote{\url{https://www.swift.ac.uk/analysis/nhtot/}}; \citealt{2013MNRAS.431..394W}) and the GRB redshift to ${\rm z}=1.3195$ \citep{2014GCN.17198....1D}. Given that the number of data bins is low compared to the number of parameters of the model, we further reduce the model complexity. We assume no temporal evolution for the host galaxy dust and total hydrogen absorption, and we also pair the optical and X-ray spectral indexes of both SEDs. Additionally, we assume an Small Magellanic Cloud (SMC) profile for the host galaxy dust absorption \citep{2007MNRAS.377..273S}. For the broken power law model, we also link the optical and the X-rays indexes as $\beta_{\rm xrays} = \beta_{\rm opt} + 0.5$, following closure relations for slow cooling synchrotron spectrum with the cooling break between the optical and X-rays bands (e.g., \citealt{1998ApJ...497L..17S,2013NewAR..57..141G}). We leave the energy break ($E_{\rm break}$) as a free parameter constrained between the optical and the X-ray bands. The data are best fitted with a broken power law model with $\chi^2/$dof$ \, = 21.2/17$ and $p$-value$ \,=0.2$; in comparison, a power law model yields worse $\chi^2$ statistics ($\chi^2/$dof$ =  39.5/ 19$ and $p$-value$=0.004$). The  best-fitting parameters for the broken power law model are a spectral index $\beta_{\rm opt}=0.6 \pm 0.2$ (i.e. $\beta_{\rm xrays}=1.1 \pm 0.2$), energy breaks at $E_{\rm break, 1} = 0.08 ^{+0.04} _{-0.03} \,$keV and $E_{\rm break, 2} = 0.015 \pm 0.005\,$keV (see Fig.~\ref{fig:SED_broadband}). In the host galaxy rest frame, the reddening is $E_{ B-V, \, {\rm HG}} = 0.24 \pm 0.06$ (i.e. absorption of $A_{V, \, {\rm HG}} = 0.71 \pm 0.15 \, $mag) and the total hydrogen absorption is $N_{\rm H, \, HG}= (3 \pm 2) \times 10^{21}\,$cm$^{-2}$. MW and Large Magellanic Cloud (LMC) dust profiles yield compatible results with $A_{V, \, {\rm HG}} = 0.74 \pm 0.18 \, $mag and $A_{V, \, {\rm HG}} = 0.76 \pm 0.16 \, $mag, respectively.

\section{Discussion} \label{sec:discussion}

Here, we discuss the nature of the fireball, shock properties and magnetic field structures in GRB 141220A, inferred from the measured properties of the observed light curves and optical polarization. In Section \ref{sec:FS}, we show that from the very start of our observations ---beginning $86\,$s post-GRB---  the afterglow properties are consistent with those exhibited by classical forward shocks observed in optical afterglows at late times ---hours to days after a GRB. In Section \ref{sec:polarization_discuss}, we discuss all possible contributions to the low polarization of $P_{BV} = 2.8 _{-  1.6} ^{+  2.0} \, \%$ (2$\sigma$) from $129.5\,$s to $204.3\,$s post-burst and conclude that the polarization of the afterglow emission itself is very small. GRB 141220A polarization is expected from theoretical predictions but is at odds with the claim by \cite{2012ApJ...752L...6U} of polarized forward shock emission in GRB 091208B. In Section \ref{sec:sample},  we compare the multiwavelength properties of GRB 141220A and GRB 091208B, and put their properties into a wider context of other GRBs with dominant forward shocks emission and optical polarization data in the first hour after the burst.

\subsection{A Classical Forward Shock at Early-Times} \label{sec:FS}

The optical emission of GRB 141220A decays with a temporal index $\alpha _{\rm opt}= 1.091\pm 0.008 $ in the {\it BV/r}-equivalent{\it /R/I} bands during LT observations ($86-2200 \,$s post-burst), which is in agreement with $\alpha _{r} = 0.96 \pm 0.11$ reported from GTC observations at $20.6-38.7\,$min post-burst \citep{2014GCN.17198....1D}. There is also no evidence of additional spectral components in the optical as there is no colour evolution through RINGO3 bands (see Fig.~\ref{fig:LC_GRB141220A_fit} bottom panel).

The {\it Swift} XRT X-ray emission can be modelled with a broken power law\footnote{A power law fit presents worse
$\chi^2$ statistics: $\chi^2$/dof= 51.8/26 and {\it p}-value= 0.002.} with $\chi^2/{\rm dof}=14.8/25$ and $p$-value=0.9. It consists of plateau-like emission ($\alpha _{\rm xrays, 1}=-0.25 \pm 0.18$) followed by a steepening of the light curve ($\alpha _{\rm xrays, 2} = 1.39\pm 0.08$) at $214^{+16} _{-29}\, $s post-burst, which we identify as the normal afterglow decay phase \citep{2006ApJ...642..389N}.

The optical and the X-rays domains satisfy closure relations for interstellar medium (ISM) profile, slow cooling regime and with the cooling break frequency ($\nu_{\rm c}$) laying in between the bands ($\nu_{\rm opt} < \nu_{\rm c} < \nu_{\rm xrays}$; \citealt{1998ApJ...497L..17S,2013NewAR..57..141G}). The expected decay rate difference $\Delta \alpha = \alpha _{\rm xrays} - \alpha _{\rm opt} = 0.25$ is consistent with the need of a spectral break in the SED modelling (Section~\ref{sec:spectra_sed}). The cooling frequency is expected to evolve like $\nu_{\rm c} \propto t^{-\alpha_{\rm c}}$ with $\alpha_{\rm c} = 0.5$, which is compatible with $\alpha = 0.7 \pm 0.4$ derived from the two SEDs epochs. In closure relations conditions, the optical decay $\alpha _{\rm opt}= 1.091 \pm 0.008$ corresponds to a typical electron index $p_{\rm CR} =2.45$ (e.g., \citealt{2008ApJ...682L...5S,2010ApJ...716L.135C}), a X-rays temporal decay of $\alpha_{\rm xrays,CR} = 1.34$ and spectral indexes $\beta_{\rm opt, CR}=0.73$ and $\beta_{\rm xrays, CR}=1.23$. These are in agreement with the ones measured: $\alpha _{\rm xrays,2} = 1.39\pm 0.08$ and $\beta_{\rm opt}=0.6 \pm 0.2$ (i.e. $\beta_{\rm xrays}=1.1 \pm 0.2$).

\begin{figure*}
  \centering
  \includegraphics[width=\textwidth]{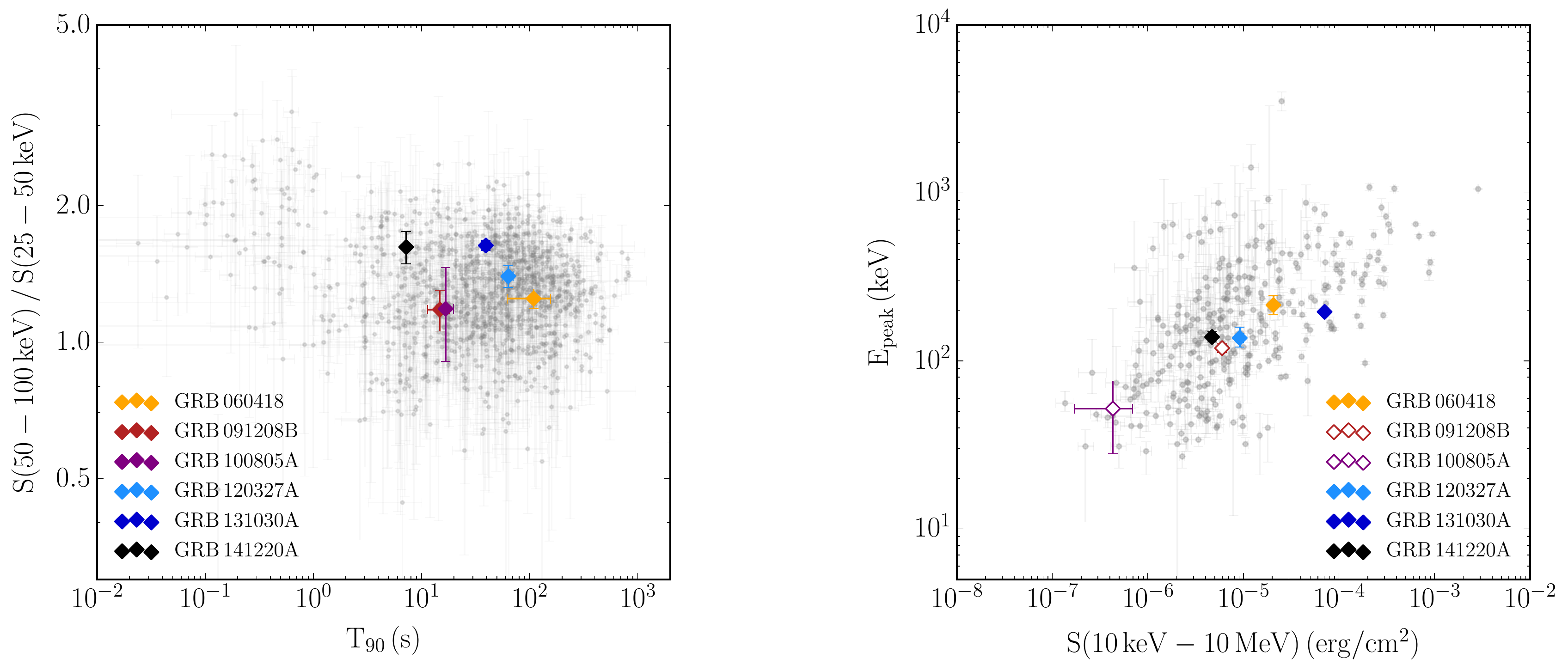}
     \caption{{\it (Left)} In grey, the GRBs duration (T$_{90}$) and the hardness ratio of {\it Swift} BAT sample \citep{1993ApJ...413L.101K,2016ApJ...829....7L}. {\it (Right)} In grey, the GRBs fluence in the $10\,$keV-$10\,$MeV band and the peak energy (E$_{\rm peak}$) of Konus-{\it Wind} sample \citep{2017ApJ...850..161T,2021ApJ...908...83T}. In colour, GRBs with early-time polarization measurements and significant forward shock contribution. We note that GRB 091208B fluence corresponds to the 8$ \,$keV-35$ \,$MeV energy range of {\it Fermi} GBM \citep{2012MNRAS.421.1256N}; GRB 100805A fluence and E$_{\rm peak}$ are derived from the $15-350 \,$keV {\it Swift} BAT time-averaged spectrum and constrained fitting a Band function with the low-energy photon index fixed to $\alpha=-1$ and the high-energy photon index to $\beta  < -2$ ($\chi^2$/dof=11.7/9 and {\it p}-value=0.2). }
    \label{fig:Swift}
\end{figure*}

\subsection{What Can Contribute to the Polarization of Early-Time Forward Shocks?} \label{sec:polarization_discuss}

Forward shock emission was predicted theoretically to be unpolarized \citep{1999ApJ...526..697M}, as the external shock sweeps up circumburst material containing weak, tangled magnetic fields. This prediction was confirmed observationally with detections of low polarization $P \sim 1\% - 3\%$ in late-time afterglow measurements made hours to days after the burst \citep{1999Sci...283.2073H,1999A&A...348L...1C,1999ApJ...523L..33W,2000ApJ...544..707R}. These low polarizations were not thought to be intrinsic to the afterglow emission but to be associated with ISM dust or jet geometry effects \citep{2004ASPC..312..169C}. However, the early-time polarization detection of $P=10.4\% \pm 2.5\%$ at $149-706\,$s post-burst for GRB 091208B forward shock started a debate on the intrinsic polarization of forward shocks during the first $\sim 10^{3} \,$s after the burst \citep{2012ApJ...752L...6U}. In contrast, our observations of GRB 141220A forward shock measure low polarization levels at early-times: $P_{BV} = 2.8 _{-  1.6} ^{+  2.0} \, \%$ (2$\sigma$) from $129.5\,s$ to $204.3\,$s post-burst and $P_{BV}<5\%$ (2$\sigma$) up to $726\,$s post-burst, which are comparable to late-time studies.
 
GRB 141220A is interesting for polarization observations at early times due to its brightness and a light curve decay typical of late-time forward shocks ($\alpha_{\rm opt} \sim 1$; e.g. \citealt{1997ApJ...485L...5W}). The optical emission does not present statistically significant "bumps" at early times (within $3\sigma$) ---usually associated with inhomogeneities in the jet or the ISM \citep{2002A&A...396L...5L,2003ApJ...598..400N} ---which reduces any issues regarding a possible break of the fireball symmetry that could introduce additional polarization signals \citep{2003A&A...410..823L, 2003ApJ...594L..83G,2004ApJ...602L..97N}. We now discuss possible mechanisms that could produce the observed polarization at very early times.

\subsubsection{Polarization Caused by the Jet Geometry} \label{sec:geometry_discuss}

Due to the relativistic beaming effect, only a small region of the physical size of the jet is visible ($1/\Gamma <  \theta_{\rm j}$). If the forward shock emission contains tangled magnetic fields in the plane of the shock with a coherence length of the order of plasma skin depth scales, the polarization is cancelled out for a face-on jet \citep{1999ApJ...526..697M}. However, high polarization is expected when the spherical symmetry is broken. For a homogeneous jet, the polarization peaks before and after the jet break \citep{1999ApJ...524L..43S,1999MNRAS.309L...7G} ---when the ejecta has slowed down to $1/\Gamma \sim \theta_{\rm j}$ (when on-axis) and the observer notices the edge of the jet (see also \cite{2004MNRAS.354...86R} for the case of structured jets).

For GRB 141220A, there are no signs of a steepening of the X-rays light curve up to $\sim 3 \times 10^4 \, $s post-burst. It is highly unlikely that we are detecting net polarization from a jet break during the first minutes after the explosion \citep{2009ApJ...698...43R} ---requiring very small electron index. Additionally, edge-on/off-axis GRBs are predicted to have softer prompt emission, lower fluence and $E_{\rm peak}$ (e.g., \citealt{2002ApJ...571L..31Y,2003ApJ...594L..79Y,2004ApJ...607L.103Y, 2019MNRAS.482.5430B,2020MNRAS.491.3343G}); GRB 141220A is on the average of Konus-{\it Wind} sample (see Fig.~\ref{fig:Swift} right; \citealt{2017ApJ...850..161T,2021ApJ...908...83T}).

\subsubsection{Post-shock Induced Polarization: The GRB Environment} \label{sec:dust_discuss}

ISM dust can induce non-negligible levels of polarization to an unpolarized source via dichroic absorption \citep{1975ApJ...196..261S}. The light is preferably absorbed when the electric field is parallel to the long axis of the dust grains. Because the short axis aligns with the magnetic field, the polarization traces the projected direction of the local magnetic field and the physical properties of dust grains. 

The wavelength-dependent behaviour of polarization was characterized with MW stars by the Serkowski empirical relation $P = P_{\rm 0} \exp \big[-K \ln^2  \big( \lambda_{\rm max} / \lambda \big)\big]$ \citep{1975ApJ...196..261S,1992ApJ...386..562W}, where $\lambda_{\rm max}({\rm \mu m})=R_{V}/5.5$ is the wavelength at which the polarization is maximum, $K = 0.01 \pm 0.05 + (1.66 \pm 0.09)\lambda_{\rm max}$ and $P_0  \lesssim 9 \, E_{ B-V}$ is an upper limit on the induced polarization that depends on the number of dust grains and the efficiency of the alignment. More recently, there has been indications that this limit is higher ($P_{\rm 0} \lesssim 13 \, E_{B-V}$) at low column densities in the MW (N$_{\rm H}<5\times10^{20}\,$cm$^{-2}$; \citealt{2020A&A...641A..12P,2019A&A...624L...8P}). Additionally, \cite{2002ApJ...564..762F} found that for the range $0.01 < E_{B-V} <1$, the average polarization dependency with extinction can be fitted with $\langle P_{0} \rangle = 3.5 \, E_{B-V}^{ \,0.8}$.

The 11.5 mag star in the {\it R/I} band field-of-view presents polarization $P_{\lbrace R, I \rbrace, \, {\rm MW}} = 0.22 _{- 0.11} ^{+ 0.13}  \%, 0.23_{- 0.14} ^{+ 0.18} \,  \%$ (2$\sigma$), which gives an estimate of the polarization induced by the Galactic ISM in the GRB 141220A line-of-sight. We note that the measured polarization agrees with the maximum expected $P_{\lbrace R, I \rbrace, \, {\rm MW}}  \lesssim 0.29 \%$ ($2\sigma$) for $E_{B-V, \, {\rm MW}} = 0.0128 \pm 0.0005$ at low column densities ($N_{\rm H, \, MW}= 1.33 \times 10^{20}\,$cm$^{-2}$). Using the Serkowski law, we determine that the polarization induced by the ISM in the {\it BV} band is $P_{BV, \, {\rm MW}}=0.23\%$. 

In addition to the polarization induced by the Galactic dust, we have the contribution induced by the propagation of the light along the host galaxy ISM. GRB 141220A is within the top $30\%$ of the most extinguished GRBs of \cite{2013MNRAS.432.1231C} sample, with $A_{V, \, {\rm  HG}} = 0.71 \pm  0.15 \,$mag (i.e. $E_{B-V, \, {\rm  HG}} = 0.24 \pm 0.06$; Section \ref{sec:spectra_sed}). To check GRB 141220A host galaxy dust contribution to the intrinsic afterglow polarization, we assume that the redshifted Serkowski law is valid for the host galaxy SMC profile\footnote{SMC and LMC polarimetric studies found that the Serkowski law was still valid ---with smaller $\lambda_{\rm max}$ on average in the SMC \citep{1983ApJ...265..194C,1996ApJ...460..313C,1997ApJ...485..618R} and polarimetric studies of M31 globular clusters measured $P_{\rm obs}  \sim [7.1,15.3] \, E_{B-V}$ with a different $K(\lambda_{\rm max})$ relationship \citep{2004AJ....127.3382C}.} with $P_0  \lesssim 9 \, E_{ B-V}$ (e.g., \citealt{2004AandA...420..899K}). Taking into account the dust-absorbed spectral slope of GRB 141220A ($\beta_{\rm opt} ^{\star}$; Section \ref{sec:spectra_opt}), we estimate that the maximum induced polarization in RINGO3 bands is\footnote{MW and LMC profiles yield similar results: $P_{BV, \, {\rm HG}}  \lesssim 1.6 \%$ and $P_{BV, \, {\rm HG}}  \lesssim 1.5 \%$, respectively.} $P_{\lbrace BV, R, I \rbrace, \, {\rm HG}}  \lesssim  1.8 \%, 2.5 \%, 2.8 \%$ ($2\sigma$), with mean polarization $ \langle P_{\lbrace BV, R, I \rbrace, \, {\rm HG}} \rangle \sim 0.6\%, 0.8\%, 1.0\%$.

As suggested in late-time \citep{2003A&A...410..823L,2004ASPC..312..169C,2012MNRAS.426....2W} and early-time afterglow polarization studies \citep{2020ApJ...892...97J}, differential light extinction from dust in the host galaxy can account for the observed polarization levels ($P_{BV} = 2.8 _{-  1.6} ^{+  2.0} \, \%$). In Sections \ref{sec:reverse_discuss} and \ref{sec:patches_discuss}, we discuss the implications of a residual polarization intrinsic to GRB 141220A afterglow.

\subsubsection{Contamination from a Polarized Reverse Shock} \label{sec:reverse_discuss}

During the early afterglow, we expect contributions from the reverse and the forward shock. GRB 141220A optical light curves present no signature of reverse shock emission and the temporal and spectral properties satisfy closure relations for forward shock emission (Section \ref{sec:FS}). The absence of a reverse shock at early times suggests that is peaking at lower frequencies \citep{2007ApJ...660..489M,2013ApJ...776..119L,2015ApJ...806..179K} or high magnetization in the jet, which suppresses the afterglow emission of the reverse shock \citep{2005ApJ...628..315Z}.
 
The expected polarization of synchrotron emission is $P_0 = (p_{\rm CR} +1)/(p_{\rm CR}  + 7/3) = 72\%$ for a homogeneous field in the slow cooling regime and for an electron power law index $p_{\rm CR} =2.45$ (inferred from closure relations in Section \ref{sec:FS}; \citealt{1979rpa..book.....R}). However, if we take into account the depolarization caused by the rotation of the polarization vector of each element in a shell due to relativistic kinematic effects, the maximum polarization of the reverse shock is reduced to $P_0 \sim 50\%$ \citep{2003ApJ...597..998L}. Observationally, we have measured polarization in the range $P =10\%-30\%$ for mildly magnetized jets (GRB 090102, \citealt{2009Natur.462..767S}; GRB 120308A, \citealt{2013Natur.504..119M}). Given the low polarization levels in GRB 141220A afterglow, we rule out significant contamination from the reverse shock.

The deceleration peak of the blastwave must have been at $t_{\rm dec} \ll 86\,$s after the GRB, well before SkyCam-Z observations, otherwise it would affect the shape of the observed light curve. From the light curve modelling\footnote{We fit the optical light curves with a reverse/forward shock model assuming ISM, slow cooling regime and the spectral configuration $\nu_{\rm m, r} \ll \nu_{\rm m, f} < \nu_{\rm opt} < \nu_{\rm c, r} \leq \nu_{\rm c, f}$.}, a $P \sim 30\%-50\%$ polarized reverse shock contributing equally at the deceleration time \citep{2007Sci...315.1822M,2007ApJ...660..489M} reproduces the measured polarization levels ($P_{BV} = 2.8 _{-  1.6} ^{+  2.0} \, \%$) for peak times at $\lesssim 20-30\,$s post-burst. We note that later deceleration times are permitted if the reverse shock is suppressed and/or less polarized than the expected \citep{2019ApJ...878L..26L,2020ApJ...892...97J}.

\subsubsection{Polarization from Collisionless Shocks} \label{sec:patches_discuss}

In matter-dominated jets, the magnetic fields do not influence the fireball dynamics but they still play a crucial role in efficient synchrotron emission and particle acceleration in collisionless shocks. The afterglow modelling usually indicates equipartition magnetic energy densities of $ \epsilon_{\rm B} \sim 10^{-6} -10^{-1}$ \citep{2002ApJ...571..779P,2015MNRAS.454.1073B,2015ApJS..219....9W}. However, for the forward shock, we are left with the magnetic field of the circumburst medium that has a few $\mu$G ($ \epsilon_{\rm B} \leq 10^{-11}$ for ISM; \citealt{1999ApJ...526..697M}). Therefore, in order to increase the magnetic energy density several orders of magnitudes, we need some mechanism to amplify the magnetic field that cannot be achieved by shock compression alone, e.g.: via plasma instabilities (e.g., Weibel instability; \citealt{1999ApJ...526..697M,2009ApJ...698L..10N}) or by macroscopic turbulences due to preshock density inhomogeneities \citep{2007ApJ...671.1858S,2011ApJ...734...77I}. In both cases, the magnetic field will have coherent length scales ($l_{\rm B}$) smaller than the observable emission area of the jet ($\theta_{\rm B} \ll 1/\Gamma <   \theta_{\rm j}$), e.g.: $l_{\rm B} \sim 10^5 \, $cm are expected for Weibel instability \citep{1999ApJ...526..697M} and $l_{\rm B} \sim 10^{13}\,$cm for macroscopic turbulences \citep{2007ApJ...671.1858S}. Magnetohydrodynamics simulations predict $P \leq 2\%$ for Richtmyer-Meshkov instability \citep{2011ApJ...734...77I}, which is in agreement with GRB 141220A forward shock polarization.

Observationally, we can estimate the coherent length scale of the magnetic field assuming a polarization model with magnetic domains \citep{1999ApJ...511..852G,2012ApJ...752L...6U}. If each patch contains ordered magnetic fields, the local polarization would be the theoretical expected for synchrotron emission when accounting for relativistic kinematic effects $P_0 \sim  50 \%$ \citep{2003ApJ...597..998L}. Because the polarization angle is randomly oriented in each magnetic domain, the total observed polarization will be reduced by the number of visible patches ($N$) as $P = P_0 / \sqrt{N}$. The observable region of the jet ($\sim R/\Gamma$) increases when the ejecta decelerates ---where $R$ is the emission radius and $\Gamma$ is the Lorentz factor the blastwave. Therefore, more patches will be visible with time, which will cause random fluctuations in the polarization angle and the polarization degree will gradually decrease. Following \cite{2012ApJ...752L...6U} derivation, we estimate the coherent length scales of the magnetic field as $ l_{\rm B} \sim R/(\Gamma \sqrt{N})$ given that $\Gamma \sim 260 \, E_{0}^{1/8} \, n_0 ^{-1/8} \,t_{\rm rest} ^{-3/8}$ and $R \sim 3.2 \times 10^{16} \, E_{0}  ^{1/4} \, n_{0} ^{-1/4} \, t_{\rm rest} ^{1/4} \, $ \citep{1997ApJ...489L..37S}. Consequently, 

\begin{equation}
l_{\rm B} \sim 10^{14} \, \bigg(\frac{E_{\rm 0} }{10^{52} \, {\rm  erg}}\bigg) ^{1/8} \, \bigg(\frac{n_{0}}{1  \, {\rm cm^{-3}}}\bigg) ^{-1/8} \, \bigg(\frac{t_{\rm rest}}{1\, {\rm s}} \bigg)^{5/8} \, P \,  P_0 ^{-1} \, {\rm cm}  
\label{eq:length}
\end{equation} where $E_{\rm 0}$ is the isotropic-equivalent kinetic energy, $n_0$ the circumburst medium density and $t_{\rm rest}$ is the rest-frame time after the burst. $E_{\rm 0}$ is related to the isotropic energy released in $\gamma$-rays as $E_{\rm iso} = \eta E_{\rm 0}$; we adopt $\eta= 0.2 $ for the radiative efficiency parameter. If we assume that the observed polarization $P_{BV} = 2.8 _{-  1.6} ^{+  2.0} \, \%$ is intrinsic to the forward shock and $E_{\rm iso} =(2.29 \pm 0.12) \times 10^{52}\, $erg \citep{2017ApJ...850..161T}, the coherent length scale of the magnetic field is $l_{\rm B} \sim 10^{14} \, n_{0} ^{-1/8} \,$cm.

\subsection{Polarization Properties of Forward Shocks} \label{sec:sample}

 \begin{figure}
	\includegraphics[width=\columnwidth]{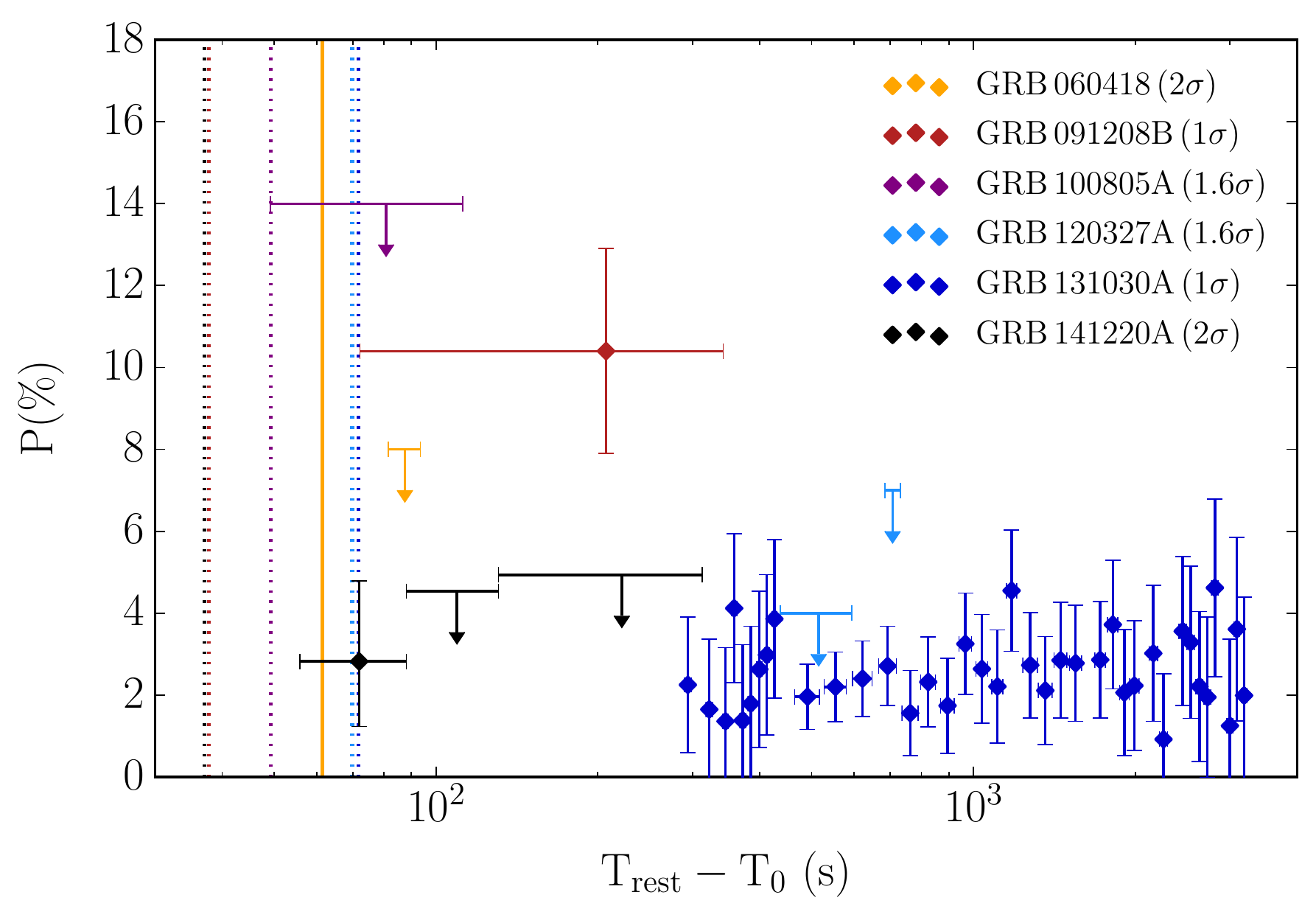}
    \caption{Rest-frame polarization observations in the optical of GRBs with light curves interpreted as early-time forward shocks. The vertical dotted lines correspond to the upper limits of the deceleration time (the earliest optical detection). The solid vertical line of GRB 060418 is the peak time of the fireball deceleration. {\it References:} GRB 060418 \citep{2007Sci...315.1822M,2007AandA...469L..13M}, GRB 091208B \citep{2009GCN.10260....1N,2012ApJ...752L...6U}, GRB 100805A \citep{2017ApJ...843..143S}, GRB 120327A \citep{2017ApJ...843..143S, 2017AandA...607A..29M} and GRB 131030A \citep{2013GCN.15414....1B,2014MNRAS.445L.114K}.}
    \label{fig:Polarization_all}
\end{figure}

\begin{table*}
	\centering
		\caption{Rest-frame properties of GRBs with optical polarization measurements ($P_{\rm obs}$) at early times ($t_{\rm obs, rest}$) and afterglow optical decays ($\alpha_{\rm opt}$) interpreted as forward shock emission.
Note that we only quote the first polarization measurement after the burst. We also state the rest-frame dust absorption for the host galaxy (A$_{\rm V, HG, rest}$) and the MW (A$_{\rm V, MW}$; \citealt{1998ApJ...500..525S}) and, the total hydrogen absorption for the rest-frame host galaxy (N$_{\rm H, \, HG, \, rest}$) and the MW (N$_{\rm H, \,  MW}$; \citealt{2013MNRAS.431..394W}). We list the duration of the bursts in the 15-350 keV {\it Swift} BAT band (T$_{\rm 90, rest}$; \citealt{2016ApJ...829....7L}), the peak energy ($E_{\rm peak, rest}$), the isotropic energy (E$_{\rm iso}$), the deceleration time of the fireball ($t_{\rm dec, rest}$) and the Lorentz factor at the deceleration time for ISM in thin shell regime ($\Gamma_{\rm dec}$; \citealt{1999ApJ...520..641S}).}
	\label{tab:summary_GRBspol}
	\begin{tabular}{c  c cc  cc cc cc c}
		\hline
		GRB &  z & $t_{\rm obs, rest}$ & $P_{\rm obs}$ &  $\alpha_{\rm opt}$  & A$_{\rm V, MW}$ & A$_{\rm V, HG, rest}$  & N$_{\rm H, \,  MW}$ & N$_{\rm H, \, HG, \, rest}$ \\
		& & (s) & ($\%$) & & & & ($10^{21}\, $cm$^{-2}$) & ($10^{21}\, $cm$^{-2}$) \\
		\hline
		060418 &  1.49 & $81.5-93.6$   & $<8$ & $1.28 \pm 0.05$  & 0.70 & $0.17 \pm 0.02$ & 1.59  & $2.80 ^{+0.97} _{-0.85}$ \\
		091208B & 1.063 & $72.2-342.2$   & $10.4 \pm 2.5$ & $0.75\pm 0.02$ & 0.16 & $0.95^{+0.22}_{-0.20}$ & 0.58  & $7.81 ^{+1.40} _{-1.21}$ \\
		100805A &  1.85  & $49.2-112.3$  & $<14$ & $0.86\pm 0.04$ &  0.58 & - & 2.64 & $4 ^{+6} _{-4}$ $^a$ \\
		120327A & 2.81 & $436.7-594.0$ & $<4$& $1.22 \pm 0.02$ & 1.04 & $<0.03$ & 2.66 & $10 \pm 2$\\
		131030A  & 1.294 & $289.9-298.6$ & $2.25\pm 1.65$ & $0.78 \pm 0.02$ & 0.18 & $0.01^{+0.03} _{-0.05}$ & 0.56 & $4.6 ^{+1.5} _{-1.3}$ \\
		141220A & 1.3195 & $55.9-88.1$   & $2.8 ^{+2.0} _{-1.6}$ &  $1.091 \pm 0.008$ & 0.04 & $0.71 \pm 0.15$ & 0.13 & $3 \pm 2$\\
		\hline
	\end{tabular}
	\begin{tabular}{c cc ccc}
		\hline
		GRB & $T_{\rm 90, rest}$& $E_{\rm peak, rest}$ & E$_{\rm iso}$ &  $t_{\rm dec, rest}$ & $\Gamma_{\rm dec}$ \\
		 & (s) & (keV) &($10^{52}$ erg) & (s) &  \\
		\hline
		060418  & $44 \pm 19$ & $535 _{-62} ^{+77}$ & $14.7 \pm 0.9$ & $61 \pm 4$ & 198  \\
		091208B  & $7.2 \pm 1.6$ &  $246 \pm 15 $ & $1.97 \pm 0.06$  & $< 37.8$ & >184  \\
		100805A & $5.8\pm 0.7$ & $148 \pm 68 $ $^a$ & $1.2 _{-0.5} ^{+0.8}$ $^a$ & $ <49.2$ & >157  \\
		120327A  & $16.7 \pm 1.8$ & $522 _{-61} ^{+84}$ & $19.1 _{-1.6} ^{+ 1.9}$ & $< 69.8$ &  >195 \\
		131030A & $17.2 \pm  1.6$ & $450 \pm 14$ & $32.7 \pm 1.3$ & $< 71.7$ &  >206  \\
		141220A  & $3.1 \pm  0.2$ & $322 _{-21} ^{+23}$ & $2.29\pm 0.12$ & $< 37.1$ &  >189 \\
		\hline
	\end{tabular}
    \begin{tablenotes}
        \item[a] {\it References:} GRB 060418 \citep{2006GCN..4969....1D,2007Sci...315.1822M,2007MNRAS.377..273S,2007AandA...469L..13M, 2021ApJ...908...83T}, GRB 091208B \citep{2009GCN.10263....1W,2009GCN.10260....1N,2012ApJ...752L...6U,2012A&A...537A..15S,2012MNRAS.421.1256N}, GRB 100805A \citep{2012MNRAS.426L..86O, 2017ApJ...843..143S}, GRB 120327A \citep{2012GCN.13133....1P,2014A&A...564A..38D,2017ApJ...843..143S,2017AandA...607A..29M, 2021ApJ...908...83T}, GRB 131030A \citep{2013GCN.15414....1B,2013GCN.15408....1D,2013GCN.15407....1X,2014MNRAS.445L.114K,2015MNRAS.449.2919L,2017ApJ...850..161T} and GRB 141220A (\citealt{2014GCN.17198....1D,2017ApJ...850..161T}; {\it this work}).
         \item[a] $^a$ GRB 100805A N$_{\rm H, \, HG, \, rest}$ is derived from {\it Swift} XRT $0.3-10 \,$keV spectra and the E$_{\rm peak}$ and E$_{\rm iso}$ are computed from {\it Swift} BAT $15-350 \,$keV spectra.
    \end{tablenotes}
\end{table*}

Very early-time follow-up is still rare and dominated by small robotic telescopes, hence there is only a small sample of bright optical afterglows with polarization measurements (e.g., \citealt{2017ApJ...843..143S}). For those GRBs with polarization measurements, optical light curves consistent with a power-law decay and with the emission interpreted as forward shock emission (at least $50\%$ of the total photons), we summarize their rest-frame properties in Fig.~\ref{fig:Polarization_all} and Table \ref{tab:summary_GRBspol}.

For all GRBs in Table \ref{tab:summary_GRBspol} sample, the polarization observations are well after the end of prompt emission ($t_{\rm obs} /T_{\rm 90} > 1$) and no contamination from prompt photons is expected (e.g., GRB 160625B, \citealt{2017Natur.547..425T}; GRB 190114C, \citealt{2020ApJ...892...97J}). We note that GRB 120327A optical light curve has a similar decay to GRB 141220A with $\alpha_{\rm opt} \sim 1.2$ \citep{2017ApJ...843..143S}, GRB 091208B/GRB 100805A/GRB 131030A have shallower light curves with $\alpha_{\rm opt} \sim 0.8$ \citep{2012ApJ...752L...6U,2014MNRAS.445L.114K, 2017ApJ...843..143S} and GRB 060418 measurement was taken during the broad peak of the fireball deceleration (decaying later with $\alpha_{\rm opt} \sim 1.3$; \citealt{2007Sci...315.1822M}). Table \ref{tab:summary_GRBspol} GRBs are characterized as long in the duration-hardness clustering of {\it Swift} BAT sample \citep{1993ApJ...413L.101K,2016ApJ...829....7L} ---with GRB 141220A being the shortest and one of the hardest (see Fig.~\ref{fig:Swift} left). GRB 141220A fluence and peak energy is on the average of the {\it Swift} BAT sample (see Fig.~\ref{fig:Swift} right). They also follow the Amati relation (E$_{\rm iso}$-E$_{\rm peak, rest}$ correlation; \citealt{2002A&A...390...81A}) within $1\sigma$ and $2\sigma$ of the scatter around the best-fitting line of \cite{2021ApJ...908...83T} and \cite{2012MNRAS.421.1256N}, respectively. To derive the bulk Lorentz factor ($\Gamma_{\rm dec}$; Table \ref {tab:summary_GRBspol}), we assume ISM profile and thin shell regime and we use \cite{1999ApJ...520..641S} derivation

\begin{equation}
\Gamma_{\rm dec} = \Bigg( \frac{3 E_{\rm 0} }{32 \pi m_{p} c^5 n_0 } \Bigg) ^{1/8} \, t_{\rm dec, rest} ^{-3/8} \,
\end{equation}where $t_{\rm dec, rest}$ is the deceleration time of the fireball and $m_{p}$ is the proton mass. The $E_{\rm iso}$, $E_{\rm peak, rest}$ and $\Gamma_{\rm dec}$ limits of Table \ref{tab:summary_GRBspol} GRBs are compatible within $1\sigma$ of \cite{2018A&A...609A.112G} sample median.

GRB 120327A \citep{2017ApJ...843..143S} and GRB 131030A \citep{2014MNRAS.445L.114K} measured low values of polarization for forward shock emission, similar to our early-time result for GRB 141220A  (see Fig.~\ref{fig:Polarization_all}). Furthermore, \cite{2014MNRAS.445L.114K} also interpreted the polarization measurements to be dust-induced (from the MW). However, GRB 091208B \citep{2012ApJ...752L...6U} polarization is significantly higher than for those GRBs. GRB 091208B and GRB 141220A have similar high-energy properties (E$_{\rm iso}$, E$_{\rm peak}$ and fluence) but different polarization and afterglow decays (i.e. electron index; see Table \ref{tab:summary_GRBspol}).  \cite{2012ApJ...752L...6U} measured $P=10.4\% \pm 2.5\%$ for GRB 091208B forward shock at a rest-frame time $t_{\rm rest} \sim 207\,$s, which corresponds to $l_{\rm B, grb09} \sim 7 \, l_{\rm B, grb14} \,\, (n_{\rm 0, grb09}/n_{\rm 0, grb14}) ^{-1/8} $ (see Eq.~\ref{eq:length}). This magnetic field length scale predicts polarization $P \sim 20\%$ for GRB 141220A at the time of observations ---far in excess of that measured. Therefore, given that GRB 091208B and GRB 141220A have similar E$_{\rm iso}$, either the magnetic field length scale is almost an order of magnitude smaller in GRB 141220A or the circumburst medium density is $10^6$ times greater in GRB 091208B ---such that \cite{2012ApJ...752L...6U} observed a smaller emitting region. However, we note that this would result in significant differences in both afterglows. Without the fine-tuning of the microphysical parameters in
  \begin{equation}  \label{eq:vc}
 \begin{aligned}
\nu_{\rm c,f} = {\rm  6.3 \times 10^{15} \, n_0 ^{-1} \,  (1+ z)^{-1/2}  \, (1+ Y)^{-2}  \,  \bigg(\frac{\epsilon_B}{10^{-2}} \bigg)^{-3/2}} \\
\times  {\rm  \bigg(\frac{E_{0}}{10^{52} erg} \bigg)^{-1/2} \,  \bigg(\frac{t}{day} \bigg)^{-1/2} Hz}
 \end{aligned}
\end{equation} and ignoring Synchrotron Self Compton effects (Y is the Inverse Compton parameter; e.g. \citealt{2007ApJ...655..989Z}), the cooling frequency would be already under the optical band at the start of GRB 091208B observations or over the X-ray band for GRB 141220A ---inconsistent with the optical and X-ray data of both GRBs. Additionally, GRB 091208B high polarization was measured from a wide time-window and it is an average of the time-evolving polarization vector ---favouring a more stable large-scale magnetic field component comparable to the reverse shock in GRB 090102 \citep{2009Natur.462..767S}. This suggests that GRB 091208B afterglow could have contribution from a polarized reverse shock component (e.g., Section \ref{sec:reverse_discuss}), be the emission of a long-lasting polarized reverse shock \citep{2007ApJ...665L..93U} or less likely ---because the light curve is consistent with ISM profile--- be embedded in the magnetized wind from the progenitor \citep{2003ApJ...594L..83G}.

\section{Conclusions} \label{sec:conclusions}

We used the 2-m autonomous robotic optical Liverpool Telescope to measure the early optical afterglow of {\it Swift} GRB141220A and its multi-colour polarization. The polarization is already low $P_{BV} = 2.8 _{-  1.6} ^{+  2.0} \, \%$ (2$\sigma$) at $129.5-204.3 \,$s after the burst (i.e. $55.9-88.1\,$s in the GRB rest-frame) in the RINGO3 {\it BV} band, and continues to be low  $P < 5\%$ (2$\sigma$) to $727\,$s post-burst.

The temporal and spectral properties of the afterglow emission are consistent with a classical forward shock. We exclude alternative interpretations: a special viewing angle to an off-axis jet, which would require an anomalously low electron index, or an edge-on jet, for which achromatic breaks would be expected but are not observed.

Extinction through the Milky Way along the line-of-sight to GRB 141220A is low; the Galactic dust is expected to induce only $P \sim 0.2\%$ polarization. We investigate different mechanisms that can contribute to the total polarization budget ($P_{BV} = 2.8 _{-  1.6} ^{+  2.0} \, \%$):

\begin{enumerate}

\item The obscuration in the GRB host galaxy is high; we estimate that host galaxy dust could induce up to $P \lesssim  1.8 \%$ at most ---consistent with our observations.

\item Reverse shocks are usually only detectable in the first minutes after the burst if bright ---their afterglows have distinct temporal properties and fade more quickly than the slowly emerging forward shock radiation. Theoretical predictions that reverse shocks should be polarized ($P \sim  50\%$) has been confirmed observationally in mildly magnetized GRBs ($P=10\%-30\%$). A reverse shock tail at the time of GRB 141220A polarization observations is very likely.

\item Forward shocks are theoretically predicted to be unpolarized or very lowly polarized with $P \leq 2\%$; late-time observations of forward shocks  ---when any primordial field has dissipated hours to a day after the burst--- and GRB 141220A early-time observations confirm this.

\end{enumerate}

The nature of very early-time polarization properties of forward shocks, however, remains debated. \cite{2012ApJ...752L...6U} detected $P=10.4 \% \pm 2.5\%$ in the first $\sim 2.5-11.8\,$minutes post-burst in GRB 091208B. They interpreted their result as the first detection of significant polarization from a forward shock. The high energy properties (E$_{\rm iso}$, E$_{\rm peak}$ and fluence) of GRB 141220A and GRB 091208B are similar, so a direct comparison of their properties is valuable. The magnetic field length scale of GRB 091208B forward shock implies $P \sim 20\%$ polarization at the time of GRB 141220A observations, which does not agree with our observations. We find that it is not likely that GRB 091208B and GRB 141220A external mediums have extreme densities and we suggest larger coherent length scales for the magnetic field of GRB 091208B, due to the presence of a large-scale magnetic field from the reverse shock.

Overall, this leads to the conclusion that forward shocks in GRBs are $P \leq 2\%$ polarized in the first few minutes after the burst and coherence scales of the generated magnetic fields are consistent with theoretical predictions. Therefore, any significant polarization measurement at early times suggests reverse shock (or prompt) emission origin.

\section*{Acknowledgements}

We thank the anonymous referee for the constructive comments that improved the paper. The Liverpool Telescope is operated on the island of La Palma by Liverpool John Moores University in the Spanish Observatorio del Roque de los Muchachos of the Instituto de Astrofisica de Canarias with financial support from the UK Science and Technology Facilities Council. This work made use of data supplied by the UK Swift Science Data Centre at the University of Leicester. The research leading to these results has received funding from the European Union's Horizon 2020 Programme under the AHEAD project (grant agreement 654215). N.J. and C.G.M. acknowledge financial support from Mr Jim Sherwin and Mrs Hiroko Sherwin. C.G.M. acknowledges support from the Science and Technology Facilities Council and the UK Research and Innovation (ST/N001265/1). A.G. acknowledges the financial support from the Slovenian Research Agency (grants P1-0031, I0-0033, J1-8136, J1-2460).

\section*{Data Availability}

The data underlying this article are available in the article and in its online supplementary material.




\bibliographystyle{mnras}
\bibliography{mnras_biblio} 

\bsp	
\label{lastpage}
\end{document}